\documentclass[12pt]{iopart}
\usepackage{graphicx}
\usepackage{xcolor}
\usepackage{array}
\usepackage{iopams}
\expandafter\let\csname equation*\endcsname\relax
\expandafter\let\csname endequation*\endcsname\relax
\usepackage{amsmath}

\usepackage{iopams}  

\begin{document}

\title[Modeling the electronic structure of organic materials]{Modeling the electronic structure of organic materials: A solid-state physicist's perspective}

\author{Caterina Cocchi,$^{1,2}$ Michele Guerrini,$^{1,2}$ Jannis Krumland,$^2$ Ngoc Trung Nguyen,$^1$ and Ana M. Valencia$^{1,2}$}
\address{$^1$ Institute of Physics, Carl von Ossietzky Universit\"at Oldenburg, 26129 Oldenburg, Germany}
\address{$^2$ Physics Department and IRIS Adlershof, Humboldt-Universit\"at zu Berlin, 12489 Berlin, Germany}
\ead{caterina.cocchi@uni-oldenburg.de}

\begin{abstract}
Modeling the electronic and optical properties of organic semiconductors remains a challenge for theory, despite the remarkable progress achieved in the last three decades. The complexity of these systems, including structural (dis)order and the still debated doping mechanisms, has been engaging theorists with different background. Regardless of the common interest across the various communities active in this field, these efforts have not led so far to a truly interdisciplinary research.
In the attempt to move further in this direction, we present our perspective as solid-state theorists for the study of molecular materials in different states of matter, ranging from gas-phase compounds to crystalline samples. Considering exemplary systems belonging to the well-known families of oligo-acenes and -thiophenes, we provide a quantitative description of electronic properties and optical excitations obtained with state-of-the-art first-principles methods such as density-functional theory and many-body perturbation theory. Simulating the systems as gas-phase molecules, clusters, and periodic lattices, we are able to identify short- and long-range effects in their electronic structure. While the latter are usually dominant in organic crystals, the former play an important role, too, especially in the case of donor/accepetor complexes. To mitigate the numerical complexity of fully atomistic calculations on organic crystals, we demonstrate the viability of implicit schemes to evaluate band gaps of molecules embedded in isotropic and even anisotropic environments, in quantitative agreement with experiments. In the context of doped organic semiconductors, we show how the crystalline packing enhances the favorable characteristics of these systems for opto-electronic applications. The counter-intuitive behavior predicted for their electronic and optical properties is deciphered with the aid of a tight-binding model, which represents a connection to the most common approaches to evaluate transport properties in these materials. 

\end{abstract}

\newpage

\section{Introduction}
Organic semiconductors are an established class of materials~\cite{brue05book} that have received considerable attention in the last three decades by a vast community across physics, chemistry, materials science, and engineering~\cite{myer-xue12}.
Since the first successes in the synthesis of ordered films of organic molecules and their integration in nanotransistors~\cite{horo98am}, the interest in organic semiconductors as opto-electronic materials has exploded. 
Their structural flexibility, chemical tunability, and low production costs have attracted hundreds of scientists and technologists to the dream of plastic electronic devices~\cite{pron-rann02pps,sing-sari06armr}.
Such a formidable experimental interest has driven the solid-state theorists' community to look systematically into the electronic and optical properties of these materials.
Pioneering results obtained from first principles date back to the last decade of the 20th century~\cite{shir-loui93prl} and to the early 2000s~\cite{ruin+02prl,buss+02apl,pusc+02prl,tiag+03prb,humm+04prl,humm-ambr05prb,humm-drax05prb2,hahn+05prb}.
These works have been largely complemented by studies employing model Hamiltonians, especially for the investigation of complex phenomena involving electron-phonon coupling~\cite{henn+99cp,casi+02prb,hoff-soos02prb,hann-bobb04prb,hann+04prb,kara-bitt04jpcb,bitt+05jcp,troi+06prl}. 
In parallel, the community active in quantum chemistry engaged itself in the investigation of organic materials by applying highly accurate methods such as coupled-cluster and configuration interaction~\cite{cram-thom01jpca,tsuz+02jacs,naka+03jpca,you+04cpl}, as well as efficient semi-empirical schemes based on the Hartree-Fock approximation~\cite{hutc+02jpca,weib-yaro02jcp,risk+04jcp}.
While both groups largely contributed to the interpretation of the experimental findings and to the progress of this research line, their efforts have remained mainly confined within their respective communities. 

In the second decade of the 21st century, the hype for organic semiconductors was substantially reduced by the rise of low-dimensional semiconductors~\cite{chho+16nrm} and halide perovskites~\cite{bren+16nrm} as novel materials for opto-electronic applications.
These emerging research areas absorbed most of the interest of the scientific community, including theorists. 
Specifically, the first-principle investigation of hybrid halide perovskites has appeared to be exceptionally challenging from the very beginning.
The accurate description of their electronic structure demands the inclusion of spin-orbit coupling~\cite{umar+14sr,fili-gius14prb,briv+14prb}.
Moreover, the complex structure of these materials combined with the peculiar softness of their lattices requires non-standard methods~\cite{bokd+16sr,mott+15natcom,he+18acsel,ghos+20jpcl,li+22jpcc}: for example, it was recently demonstrated that the lattice contribution to the screened Coulomb interaction should be explicitly incorporated in order to obtain accurate estimates of the exciton binding energies~\cite{fili+21prl}. 

In such a context, molecular semiconductor modelling was devoted to addressing more specific questions related, among others, to doping mechanisms~\cite{zhu+11cm,salz+13prl,zhu+14jpcc,mend+15ncom,salz+16acr,li+17prm,beye+19cm}, polymorphism~\cite{ambr+09njp,pith+15cgd,klet+16pccp,jaco+18mh,cocc+18pccp}, and singlet fission~\cite{zimm+11jacs,shar+13jpcl,berk+14jcp,refa+17prl,wang+18jcp,abra-mayh21jpcl,altm+22jpcl}.
Moreover, increasing attention was dedicated to disclosing the excitonic properties of organic materials~\cite{cocc+18pccp,cuda+12prb,rang+16prb,zhu+19jpcl}, including quantifying resonance energies and clarifying the character of electron-hole pairs.
These efforts, requiring an exceptionally high degree of specialization, enhanced the segregation between quantum/computational chemistry and solid-state theory. 
Both communities have relevantly contributed to address questions related to the role of the environment~\cite{sun+16jctc,cast+17cpl,li+17prm,duch+18cs} and of electron-vibrational couplings~\cite{ross-sohl09jpcc,brow+20prb,cook-bera20jcp} in the opto-electronic activity of organic materials, as well as to benchmark the performance of advanced first-principles methodologies applied to this material class~\cite{jacq+09jctc,shar+12prb,rang+16prb}.
Yet, endeavors in physics and chemistry have mostly run in parallel with very rare intersections.
Only by merging the efforts of different communities it will be possible to solve problems of increasing complexity leading to a substantial advancement in the fundamental and applied research on organic materials.

Adopting our perspective as solid-state physicists, we discuss herein the ability of density-functional theory and many-body perturbation theory, in their respective implementations for isolated and extended systems, to describe the electronic and optical properties of organic materials, ranging from isolated molecules to clusters, and crystalline solids. 
Taking anthracene as an example, we assess the role of short- and long-range effects in electronic structure and optical excitations.
Given the numerical complexity of full-fledged simulations on organic crystals, we prove the viability of implicit schemes coupled to density-functional theory to account for electrostatic interactions exerted on molecules by both isotropic and anisotropic environments, obtaining band-gap values in excellent agreement with experiments.
Finally, turning to donor/acceptor complexes, we discuss the importance of accounting for the lattice periodicity in order to describe the more favorable opto-electronic properties of co-crystals in comparison with their non-periodic counterparts.
To interpret the counter-intuitive behavior of these systems, we complement our first-principles results with a tight-binding model, thus providing a connection with established schemes to describe transport properties.

This review is organized as follows: In Section~\ref{sec:methods}, we introduce the basic formalism of density-functional theory (Subsection~\ref{ssec:DFT}) and many-body perturbation theory (Subsection~\ref{ssec:MBPT}).
Next, we discuss the electronic and optical properties of anthracene, first described atomistically in different states of matter (Section~\ref{sec:ant}) and then simulated as a single molecule embedded in various environments (Section~\ref{sec:pcm}).
Finally, we examine the physics of a well-known donor/acceptor complex formed by a $p$-doped quaterthiophene oligomer, making use of a tight-binding model to shed light on the electronic and optical behavior of the corresponding co-crystal as predicted by first-principles calculations (Section~\ref{sec:co-cry}). 
The computational details are provided in the Appendix.
 
\section{Methodology}\label{sec:methods}

\subsection{Density-functional theory for molecules and crystals: One tool for two communities}\label{ssec:DFT}

Density-functional theory (DFT) is one of the most successful methods for electronic-structure calculations.
Its theoretical foundation lies upon the Hohenberg-Kohn theorems~\cite{hohe-kohn64pr}, defining the unique correlation between the potential and the density of a many-electron system,
which can be used to represent all its ground-state properties.
The most popular and efficient recipe to implement DFT is given by the Kohn-Sham (KS) equations~\cite{kohn-sham65pr}, which map the many-body problem into an auxiliary system of non-interacting particles ruled by the Schr\"odinger equation
\begin{equation}
\left[-\dfrac{\nabla^2}{2} + v_s(\mathbf{r}) \right] \phi_{i}(\mathbf{r}) = \epsilon^{KS}_{i} \phi_{i}(\mathbf{r}).
\label{eq:KS}
\end{equation}
The eigenfunctions of Eq.~\eqref{eq:KS} are the states of the fictitious KS system of independent electrons that are used to express the electron density, 
\begin{equation}
n(\mathbf{r}) = \sum_{i=1}^{occ.} |\phi_i(\mathbf{r})|^2,
\label{eq:n}
\end{equation}
and the eigenvalues represent their corresponding single-particle energies. 
Beside the kinetic-energy term, the Hamiltonian of Eq.~\eqref{eq:KS} includes an effective potential which is given by the sum of three terms:
\begin{equation}
v_s(\mathbf{r}) = v_{ext}(\mathbf{r}) + v_H[n](\mathbf{r}) + v_{xc}[n](\mathbf{r}).
\label{eq:v_eff}
\end{equation}
The external potential, $v_{ext}$, describes the electron-nuclear Coulomb attraction, the Hartree potential, $v_H[n](\mathbf{r})$, the mean-field electron-electron coupling, and the exchange-correlation (xc) potential, $v_{xc}[n](\mathbf{r})$, includes the remaining interactions.
While the form of the first two potentials is known, unfortunately, the same is not true for $v_{xc}[n](\mathbf{r})$.
Hence, the accuracy of the results provided by the KS equations crucially depends on the approximation chosen for this quantity.

The local-density approximation (LDA), originally proposed by Kohn and Sham in their seminal paper from 1965~\cite{kohn-sham65pr}, is based on the homogeneous electron gas model and represents the lowest rank of the so-called Jacob's ladder of density-functional approximations~\cite{perd-schm01aipcp}.
The generalized gradient approximation (GGA)~\cite{perd86prb} is the next rung, with $v_{xc}$ depending on both the density and its gradients.
Higher-level approximations include, on the one hand, the so-called metaGGA functionals~\cite{tran-blah09prl,sun+15prl}, and, on the other hand, hybrid functionals, where $v_{xc}$ is enhanced by a fraction of Fock exchange~\cite{beck93jcp}.
Among the latter, global hybrids are distinguished from the range-separated ones, in which the long- and short-range parts of the Coulomb potential are augmented separately~\cite{iiku+01jcp,toul+04pra}.
Interestingly, the diffusion of these families of functionals has followed separate paths in chemistry and physics, reflecting the substantially different performance of the various approximations on (isolated) molecules and solids.
While for inorganic crystals the LDA and GGA have been extensively adopted for decades and specific flavors of range-separated hybrid functionals such as HSE~\cite{hse03} have gained popularity only recently, in the study of molecules the use of hybrid functionals has become a standard since the turn of the turn of the century.

The definition of the basis set for the KS equations (Eq.~\ref{eq:KS}) is another crucial aspect in the implementation of DFT applied to non-periodic and extended systems. 
The delocalized character of the electronic wave-functions in solids is prone to a representation in terms of pure plane-waves or to plane-waves augmented by local orbitals~\cite{gula+14jpcm}.
The solution of the KS problem in \textbf{k}-space typically exploits the crystal periodicity and enables the application of efficient parallelization schemes~\cite{gian+17jpcm}.
On the other hand, the non-periodic nature of the electronic states in molecules calls for a real-space description with localized basis sets.
Gaussian functions, atom-centered orbitals, and grid representations are the most common options ensuring an optimal trade-off between accuracy and computational costs~\cite{g16,blum+09cpc,andr+15pccp}.

\subsection{Many-body perturbation theory: Gateway to the excited states}\label{ssec:MBPT}

The efficient \textit{ab initio} scheme provided by DFT is limited by construction to the description of ground-state properties. 
In order to access excited states, it is necessary to take other routes. 
One option is to adopt the time-dependent extension of DFT (TDDFT)~\cite{rung-gros84prl}.
This approach is routinely applied both in a perturbative flavor~\cite{casi+98jcp} as well as by propagating the KS equations in real time~\cite{marq-gros04arpc}: the former approach gives access to the single-particle composition of the excitations but is limited to the linear response; the latter grants access to the response of all orders~\cite{cocc+14prl,guan+21pccp} even in presence of an external, time-dependent electric fields~\cite{degi+13chpch,krum+20jcp}, but does not provide any information about the nature of the excited states.
In both flavors, TDDFT results are crucially dependent of the choice of the exchange-correlation functional.
Calculations based on hybrid functionals are largely applied in the chemistry community to describe optical excitations of molecules~\cite{jacq+09jctc}, even in conjunction with embedding methods~\cite{neug+05jcp} to account for solvation effects. 
On the other hand, in spite of enormous efforts in the past decades to develop appropriate approximations~\cite{bott+07rpp}, TDDFT is not considered reliable to describe excitations in solids, not even when composed of molecules~\cite{sott+05ijqc,cocc-drax15prb}.
For this reason, its popularity in the solid-state physics community is relatively low.

The state-of-the-art approach to describe electronic and optical excitations in extended systems is many-body perturbation theory (MBPT)~\cite{onid+02rmp}, including the $GW$ approximation for the self-energy~\cite{hedi65pr} and the solution of the Bethe-Salpeter equation~\cite{salp-beth51pr}.
Modern implementations of MBPT~\cite{hank-sham80prb,hybe-loui85prl} are built on top of DFT in the usual frameworks outlined in Section~\ref{ssec:DFT} for both periodic (see, e.g., Refs.~\cite{marini2009,vorw+19es}) and non-periodic systems~\cite{blas+11prb,hiro+15prb,brun+16cpc}.
In this framework, the single-particle Green's function
\begin{equation}
G(\mathbf{r},\mathbf{r}',\omega) =\sum_i \dfrac{\phi_{i}^*(\mathbf{r}) \phi_{i}(\mathbf{r}')}{\omega - \epsilon^{KS}_{i} - i\eta}
\label{eq:G}
\end{equation}
is defined in terms of the non-interacting KS states and energies obtained from Eq.~\eqref{eq:KS}.
The expression of the dynamically screened Coulomb interaction
\begin{equation}
W(\mathbf{r},\mathbf{r}',\omega) = \int\text d^3r_1\, \varepsilon^{-1}(\mathbf{r},\mathbf{r}_1,\omega) \; v_C(\mathbf{r}_1,\mathbf{r}') \; ,
\label{eq:scr}
\end{equation}
in addition to the bare Coulomb potential $v_C$, includes the frequency-dependent inverse dielectric function $\varepsilon^{-1}$, which is evaluated from KS states, too~\cite{onid+02rmp}. 
Eqs.~\eqref{eq:G} and \eqref{eq:scr} are the ingredients for the construction of the self-energy in the $GW$ approximation, $\Sigma=iGW$.
In the original formulation by Lars Hedin~\cite{hedi65pr}, this equation was supposed to be solved self-consistently as part of a closed loop of equations.
In practice, the perturbative ``single-shot" approach $G_0W_0$ is typically adopted with good results especially for conventional semiconductors~\cite{rein18wircms}. 
In this framework, the quasi-particle (QP) equation dressing the KS energies with the self-energy contribution reads~\cite{arya-gurn98rpp}:
\begin{equation}
\epsilon_{i}^{QP} = \epsilon_{i}^{KS} + Z_{i} \langle\phi_i| \Re \Sigma(\epsilon_{i}^{KS}) - v_{xc} |\phi_i\rangle,
\label{eq:QP}
\end{equation}
where $Z_i$ is a renormalization factor compensating for the evaluation of $\Sigma$ at $\epsilon_{i}^{KS}$, rather than the correct $\epsilon_{i}^{QP}$, by means of a linear extrapolation.

Starting from the QP-corrected electronic structure, optical excitations are computed by solving the Bethe-Salpeter equation (BSE)~\cite{salp-beth51pr}, the equation of motion for the two-particle correlation function~\cite{stri88rnc}.
By construction, this formalism enables the description of excitons quantitatively accounting for electron-hole interactions.
In practice, the problem is expressed in terms of an effective two-particle Schr\"odinger equation
\begin{equation}
\sum_{o'u'} \hat{H}^{BSE}_{ou,o'u'} A^{\lambda}_{o'u'} = E^{\lambda} A^{\lambda}_{ou} ,
\label{eq:BSE}
\end{equation}
where $o$ and $u$ stand for occupied and unoccupied states, respectively.
In spin-unpolarized systems, the BSE Hamiltonian is expressed as:
\begin{equation}
\hat{H}^{BSE} = \hat{H}^{diag} + \hat{H}^{dir} + 2 \hat{H}^x,
\label{eq:H_BSE}
\end{equation}
where the \textit{diagonal} term, $\hat{H}^{diag}$, accounts for the energy differences between occupied and unoccupied states, $\hat{H}^{dir}$ corresponds to the \textit{direct} Coulomb attraction between the positively-charged hole and the negatively-charge electron,
\begin{equation}
\hat{H}^{dir} = - \int \text d^3r  \int \text d^3r' \, \phi_{o} (\mathbf{r}) \phi^*_{u} (\mathbf{r}') \, W(\mathbf{r},\mathbf{r}') \, \phi^*_{o'} (\mathbf{r}) \phi_{u'} (\mathbf{r}'),
\label{eq:Hdir}
\end{equation}
and the \textit{exchange} term $\hat{H}^x$ includes the \textit{exchange} electron-hole repulsion,
\begin{equation}
\hat{H}^x = \int \text d^3r \int \text d^3r'\, \phi_{o} (\mathbf{r}) \phi^*_{u} (\mathbf{r}) \, \bar{v}_{C}(\mathbf{r},\mathbf{r}') \, \phi^*_{o'} (\mathbf{r}') \phi_{u'} (\mathbf{r}').
\label{eq:Hx}
\end{equation}
Eq.~\eqref{eq:Hdir} includes the screened Coulomb interaction (Eq.~\ref{eq:scr}), with $\varepsilon$ evaluated at $\omega =0$ (static screening).
In Eq.~\eqref{eq:Hx}, $\bar{v}_C$ is the short-range part of the bare Coulomb potential accounting for local-field effects which are known to be particularly relevant in the optical response of organic materials~\cite{sott+05ijqc,cocc-drax15prb,cocc+16jcp,cocc-drax17jpcm}.

The eigenvalues of Eq.~\eqref{eq:BSE}, $E^{\lambda}$, represent excitation energies and mark the resonances in the absorption spectra.
In the context of inorganic semiconductors, where excitons are loosely bound and largely delocalized, exciton binding energies, $E_b$, are defined as the difference between excitation energies and the QP optical gap: $E_b = E^{\lambda} - E^{QP-opt}_{gap}$~\cite{rohl-loui00prb}.
In the spectra of these materials, excitons are identified by narrow resonances below the onset~\cite{fox02book}.
This definition turns out to be inappropriate for organic crystals where the optical absorption spectrum is usually formed by distinct peaks even beyond the absorption onset, which are present already in the (joint) density of states~\cite{cocc-drax15prb,cocc-drax17jcpm,guer+21jpcc}.
In this scenario, a more consistent definition of the exciton binding energy is given by $E_b = E^{\lambda}_{BSE} - E^{\lambda}_{IQPA}$, where the independent QP energies, $E^{\lambda}_{IQPA}$, are the eigenvalues of Eq.~\eqref{eq:H_BSE} with $\hat{H}^{BSE} \equiv \hat{H}^{diag}$.

The eigenvectors of Eq.~\eqref{eq:BSE}, $A^{\lambda}$, contain information about the electronic composition and the spatial distribution of the excited states.
They act as weighting factors in the transition dipole coefficients, defined as $\mathbf{t}^{\lambda}= \sum_{ou} A^{\lambda}_{ou} \langle o|\widehat{\mathbf{d}}|u\rangle$ for localized systems and in terms of the momentum operator,
\begin{equation}
\mathbf{t}^{\lambda}= \sum_{ou\mathbf{k}} A^{\lambda}_{ou\mathbf{k}} \frac{\langle o\mathbf{k}|\widehat{\mathbf{p}}|u\mathbf{k}\rangle}{\varepsilon^{QP}_{u\mathbf{k}} - \varepsilon^{QP}_{o\mathbf{k}}},
\label{eq:t}
\end{equation}
in periodic structures.
The imaginary part of the macroscopic dielectric function
\begin{equation}
\text{Im}\,\varepsilon_M=\frac{8\pi^2}{\Omega}\sum\limits_\lambda|{\mathbf{t}^{\lambda}}|^2\delta(\omega-E^\lambda),
\label{eq:Im}
\end{equation}
describes the optical absorption of the material within the unit cell volume $\Omega$.
Given the vectorial character of $\mathbf{t}^{\lambda}$, $\varepsilon_M$ is a tensor with as many non-vanishing components as enabled by the crystal symmetries.

Furthermore, the eigenvectors of Eq.~\eqref{eq:BSE} are included in the definition of hole and electron densities, defined as
\begin{equation}
\rho_{h}^{\lambda} (\textbf{r})=\sum_{ou} |A_{ou}^{\lambda}|^2 |\phi_{o}(\textbf{r})|^{2},
\label{eq:h}
\end{equation}
and
\begin{equation}
\rho_{e}^{\lambda} (\textbf{r})=\sum_{ou} |A_{ou}^{\lambda}|^2 |\phi_{u}(\textbf{r})|^{2},
\label{eq:e}
\end{equation}
respectively, for transitions between occupied states $\phi_{o}$ and unoccupied states $\phi_{u}$.
These quantities were originally introduced for isolated systems~\cite{cocc+11jpcl,deco+14jpcc}.
In solids, Eqs.~\eqref{eq:h} and \eqref{eq:e} retain the periodicity of the lattice~\cite{guer+21jpcc}: as such, they highlight the fictitiously periodic distribution of the hole and the electron but cannot provide any information about the exciton localization.
For this purpose, it is necessary to introduce the exciton wave function, a six-dimensional quantity expressed by
\begin{equation}
\Psi^{\lambda}(\mathbf{r}_h, \mathbf{r}_e) = \sum_{uo\mathbf{k}} A^{\lambda}_{uo\mathbf{k}} \phi^*_{o\mathbf{k}}(\mathbf{r}_h) \phi_{u\mathbf{k}}(\mathbf{r}_e),
\label{eq:exciton}
\end{equation}
and typically visualized by the isosurface of its square modulus with the hole or the electron position fixed.

\subsection{Polarizable Continuum Model}

The description of electronic structure provided by DFT (and even MBPT) assumes the systems to be \textit{in vacuo}.
Especially for molecules, this is hardly a realistic scenario.
Molecules are usually emedded in an environment which has a strong influence on their electronic and optical properties.
In order to capture these effects, implicit methods have been developed in quantum chemistry.
The polarizable continuum model (PCM) is one of the most popular ones and describes the surrounding medium by a single polarizable continuum interacting with the charge density of the molecule hosted in a cavity. 
The validity of this approximation depends on the system.
Since it fails to capture intermolecular interactions between static dipoles, van der Waals forces, or chemical bonding, including charge transfer, its accuracy is limited to those scenarios in which these effects are negligible, unless corresponding (usually empirical) corrections are included~\cite{toma+05cr}. 

From a mathematical point of view, the polarization response of the medium described by the PCM is included through a density-dependent interaction term added to the KS Hamiltonian, $\hat{H}^{KS}[n]$, formulated in Eq.~\eqref{eq:KS}: $\hat{H}^{KS}[n]\rightarrow \hat{H}^{KS}[n] + \hat{v}^\mathrm{PCM}[n]$.
The interaction term, $\hat{v}^\mathrm{PCM}[n]$, is a local electrostatic potential generated by the polarization of the surrounding. It can be exactly represented as arising from a surface charge density $\sigma[n]$,
\begin{align}
    v^\mathrm{PCM}[n](\textbf{r}) = \int_{\cal C} \text d^2s\,v_c(\textbf{r}-\textbf{s})\sigma[n](\textbf{s}),
\end{align}
where $v_c(\textbf{r}) = 1/|\textbf{r}|$ is the bare Coulomb interaction and $\cal{C}$ is the cavity surface, \textit{i.e.}, the interface between the molecule and the medium. In turn, $\sigma[n]$ is determined from the electrostatic potential created by the molecule's charges on $\cal{C}$:
\begin{align}
    \sigma[n](\textbf{s}) = \int_{\cal{C}}\text d^2s'\,{\cal Q}(\textbf{s},\textbf{s}')\left[\sum_{j=1}^NZ_jv_c(\textbf{s}'-\textbf{R}_j)-\int\text d^3r\,v_c(\textbf{s}'-\textbf{r})n(\textbf{r})\right],
\end{align}
where $Z_j$ and $\textbf{R}_j$ are nuclear charges and positions, respectively. The PCM matrix, $\cal{Q}$, encodes all information about the response of dielectric environment. Within the flexible integral equation formalism \cite{cances1997jcp}, it is most generally given as 
\begin{align}\label{eq:pcm_mat}
{\cal Q} = \left[\left(2\pi {\cal I}-{\cal D}_e\right){\cal S}_i+{\cal S}_e\left(2\pi {\cal I}+{\cal D}_i^*\right)\right]^{-1}\left[{\cal S}_e{\cal S}_i^{-1}\left(2\pi {\cal I}+{\cal D}_i\right)+\left(2\pi {\cal I}+{\cal D}_e\right)\right],
\end{align}
where $\cal{I}$ is the identity operator, and ${\cal S}_{i}$, ${\cal D}_{i}$ and ${\cal D}^*_i$, as well as ${\cal S}_{e}$ and ${\cal D}_{e}$ are integral operators, defined by their action on an function $u$ supported on ${\cal C}$ as~\cite{cances1998jmc}:
\begin{subequations}
\begin{align}
    \left({\cal S}_{i}u\right)(\textbf{s}) &= \int_{\cal C}\text d^2s'\, v_c(\textbf{s}-\textbf{s}')u(\textbf{s}') \label{simat.eq}\\
    \left({\cal D}_{i}u\right)(\textbf{s}) &= \int_{\cal C}\text d^2s'\left[\hat{\textbf{n}}(\textbf{s}')\cdot \nabla' v_c(\textbf{s}-\textbf{s}')\right]u(\textbf{s}') \\
    \left({\cal D}^*_iu\right)(\textbf{s}) &= \int_{\cal C}\text d^2s'\left[\hat{\textbf{n}}(\textbf{s})\cdot \nabla v_c(\textbf{s}-\textbf{s}')\right]u(\textbf{s}')\\
    \left({\cal S}_{e}u\right)(\textbf{s}) &= \int_{\cal C}\text d^2s'\, W(\textbf{s},\textbf{s}')u(\textbf{s}') \\
    \left({\cal D}_{e}u\right)(\textbf{s}) &= \varepsilon\int_{\cal C}\text d^2s'\left[\hat{\textbf{n}}(\textbf{s}')\cdot \nabla' W(\textbf{s},\textbf{s}')\right]u(\textbf{s}'),
\end{align}
\end{subequations}
where $\hat{\textbf{n}}(\textbf{s})$ is the cavity surface normal vector and $\varepsilon$ is the dielectric constant of the material directly on the exterior of ${\cal C}$. The term $W(\textbf{r},\textbf{r}')$ appearing in ${\cal S}_e$ and ${\cal D}_e$ is the screened Coulomb interaction in the external medium, disregarding the presence of the cavity. If the so-described environment is homogeneous  with dielectric constant $\varepsilon$ (a reasonable approximation for solvents or crystalline environments), it is simply given as $W(\textbf{r},\textbf{r}') = \varepsilon^{-1}v_c(\textbf{r}-\textbf{r}')$. In this case, Eq.~\eqref{eq:pcm_mat} can be simplified to 
\begin{align}
    {\cal Q} = -{\cal S}_i^{-1}\left(2\pi\frac{\varepsilon+1}{\varepsilon-1}{\cal I}-{\cal D}_i\right)^{-1}\left(2\pi{\cal I}-{\cal D}_i\right),
\end{align}
see Ref.~\cite{corni2015jpca}.
When instead the medium is anisotropic, adjustments to the formalism are needed, as discussed in Ref.~\cite{krum+21jcp} and in Section~\ref{sec:pcm} below.

\section{Organic semiconductors: From molecules to crystals} \label{sec:ant}

We start our analysis with the example of anthracene (ANT), the three-ring member of the oligoacene series.
In the following, we examine the electronic and optical properties of this system in gas and crystalline phase calculated from DFT and MBPT, and discuss the impact of the chosen modelling framework in reproducing the physics of the system.

\subsection{Electronic properties}
\label{ssec:ant-electr}
In the first step, we evaluate the frontier states of the isolated ANT molecule from DFT adopting a range-separated hybrid functional and applying the $GW$ approximation in the perturbative approach $G_{0}W_{0}$ (see computational details in the Appendix).
In Fig.~\ref{fig.ant.el}a), the isosurfaces of the highest occupied and lowest unoccupied molecular orbitals (HOMO and LUMO, respectively) are visualized in a schematic energy-level diagram including the neighboring levels, HOMO-1 and LUMO+1. 
The picture provided by these results, yielding a fundamental gap of 6.94~eV, is in excellent agreement with the experimental references for the electronic structure of this molecule (6.9~eV, see Ref.~\cite{prob-karl75pssa}).

\begin{figure}[h]
\centering
    \includegraphics[width=\textwidth]{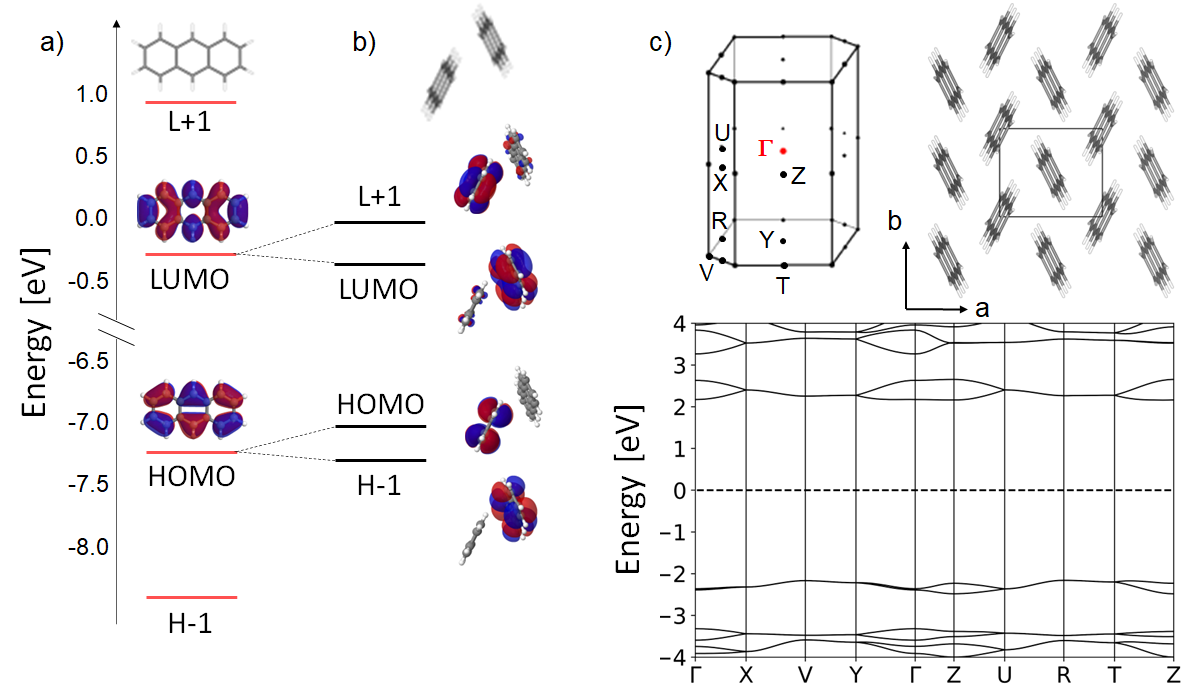}
    \caption{Energy levels and frontier orbitals of a) the isolated anthracene molecule and b) the bimolecular cluster including the two inequivalent molecules in the unit cell of the crystal. c) Crystal structure, Brillouin zone (BZ), and band structure of the anthracene crystal plotted along the path connecting the high-symmetry points highlighted in the BZ.}
    \label{fig.ant.el}
\end{figure}

In the bulk phase, ANT crystallizes in a monoclinic structure, space group $P2{_1}/a$ ~\cite{math+50ac}, including two inequivalent molecules in the unit cell, see Fig.~\ref{fig.ant.el}c), where both the real-space representation of the material and its Brillouin zone (BZ) are shown.
The resulting band structure, reported in the bottom panel of Fig.~\ref{fig.ant.el}c), exhibits the hallmarks of organic crystals.
First and foremost, the electronic states are no longer described by isolated energy levels but by bands
with a finite \textbf{k}-dispersion providing information about the carrier mobility along a certain direction.
The relatively flat character of the bands of ANT compared to their counterparts in inorganic semiconductors is related to the molecular nature of the constituents.
Since in the energy window displayed in Fig.~\ref{fig.ant.el}c) the electronic states have $\pi$/$\pi^*$ character, the dispersion is larger along those directions where the delocalization of the corresponding electronic cloud is maximized.
For this reason, in the electronic structure of longer members of the oligoacene family, such as tetracene and pentacene, the band dispersion is overall enhanced~\cite{tiag+03prb,humm-drax05prb2,cocc+18pccp,cuda+15jpcm}.
Another typical characteristic of organic crystals in the band structure of ANT is the twofold splitting of each electronic level, due to the presence of two inequivalent molecules in the unit cell.
Depending on the specific direction in reciprocal space, the ``sub-states'' are either energetically slit up or almost degenerate: the former scenario is realized when the corresponding wave function is spread over both molecules in the unit cell, thereby inducing electronic repulsion; in the latter case, the wave function is localized on only one inequivalent molecule in the unit cell, and thus the mutual repulsion is minimized~\cite{cocc+18pccp}.
As we will see in Subsection~\ref{ssec:optics}, this characteristic impacts the optical spectrum of the material giving rise to Davydov splitting~\cite{cocc+18pccp, Davydov_1964}.
The computed band-gap, 4.30~eV, indicates a substantial renormalization with respect to result obtained for the isolated molecule (6.94~eV, see Fig.~\ref{fig.ant.el}a) as a consequence of the screening exerted by the molecules forming the crystal~\cite{refa+13prb}.
Notice that the aforementioned value, 4.30~eV, overestimates the experimental band gap of the ANT crystal, ranging between 3.90 and 4.00~eV~\cite{baes+kill1973mclc,belk+grec1974pssa,riga+1977phsc}.
We can ascribe this discrepancy to the fact that the band structure reported in Fig.~\ref{fig.ant.el}c) is obtained adopting the GGA in DFT, with the QP correction to the band gap included by means of an empirical scissors operator, see details in the Appendix.
This approach is often used when experimental references are available (see, e.g., Refs.~\cite{humm+04prl,cocc+18pccp,guen+21jpcl}), given the high numerical costs of $GW$ calculations on these systems.

In order to understand whether the discussed electronic properties of the ANT crystal are primarily induced by short- or long-range effects, we investigate a non-periodic cluster formed by the two inequivalent molecules extracted from the unit cell of the crystal (see Fig.~\ref{fig.ant.el}b).
Neglecting the lattice periodicity, the energy levels are obviously dispersionless, confirming the long-range nature of this property.
However, the twofold splitting in the energy levels of the molecule is preserved and its magnitude is consistent with the result obtained in the crystal at the high-symmetry point $Z$ (see Fig.~\ref{fig.ant.el}c). This finding suggests that the short-range interactions between the molecules included in the unit cell are responsible for the Davydov splitting, hinting that this effect can be reproduced even by non-periodic models.
The isosurface plots of the orbitals shown in Fig.~\ref{fig.ant.el}b) reveal the correspondence between the HOMO and the HOMO-1 (LUMO and LUMO+1) of the cluster and the HOMO (LUMO) of the isolated molecule. 
In both manifolds, the lower-energy state is localized on the same molecule of the cluster.
The orbital segregation is consistent with the character of the corresponding states in the crystal at the $\Gamma$ point. 
The value of the fundamental gap in the cluster, 6.66 eV, is less than 0.3~eV smaller than the one computed for the isolated molecule. 
This result, in line with previous findings obtained at the same level of theory for longer oligoacenes~\cite{zeis+21jpcl}, is unsurprising and confirms that the band-gap reduction seen in the crystal cannot be captured with such a minimal model.
Yet, as discussed in Section~\ref{sec:pcm}, a convenient shortcut is available for an accurate evaluation of this quantity at affordable computational costs, without the need to simulating the full periodic crystal.

\subsection{Optical properties}\label{ssec:optics}
\begin{figure}[h!]
    \centering
    \includegraphics[width=.75\textwidth]{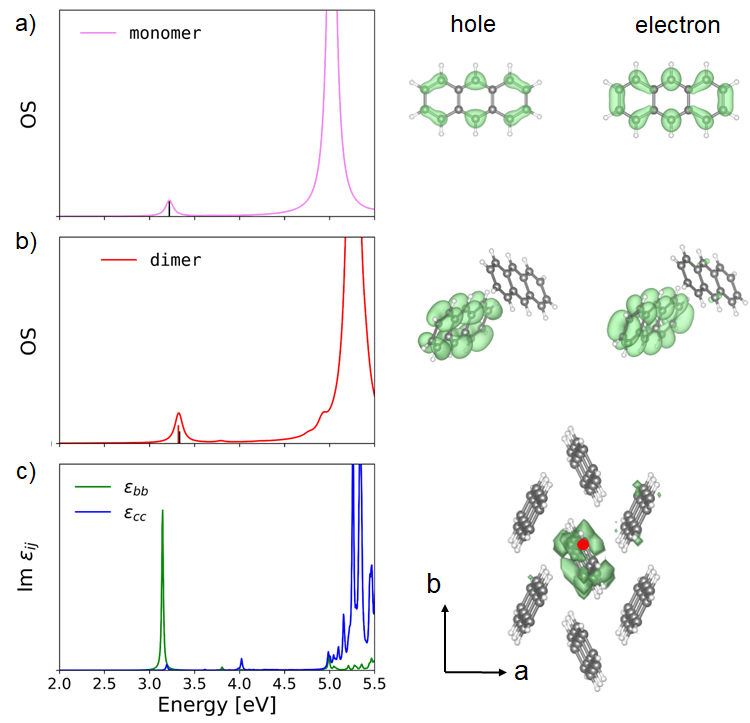}
    \caption{Optical absorption spectra (left) and electron-hole distribution (right) of a) isolated anthracene (monomer), b) a bimolecular cluster of anthracene molecules in the crystalline geometry (dimer), and c) the anthracene crystal. In panels a) and b) vertical bars mark the position of the lowest-energy resonances. In panel c), the isosurface represents the electron distribution correlated to the hole fixed at the position marked by the red dot. To plot the spectra, a Lorentzian broadening of 50~meV is used in panels a) and b), and of 10~meV in panel c).}
    \label{fig.ant.opt}
\end{figure}

We now turn to the optical properties of ANT, considering again the isolated molecule, the crystal, and the bimolecular cluster extracted from its unit cell.
The absorption spectrum calculated for gas-phase compound (Fig.~\ref{fig.ant.opt}a) is characterized by an optically active excitation at the onset. 
The energy of this resonance is in excellent agreement with the experimental value~\cite{stai+04jms} and the corresponding excitation is given by the HOMO$\rightarrow$LUMO transition polarized along the short molecular axis.
This characteristic, common to all oligoacenes~\cite{humm-ambr05prb}, can be understood from the parity of the frontier orbitals shown in Fig.~\ref{fig.ant.el}a).
Consistently, the hole and electron densities (Eq.~\ref{eq:h} and \ref{eq:e}) exhibit the same spatial distribution as the HOMO and the LUMO, respectively.
The orientation of the transition dipole moment of this excitation explains its lower oscillator strength with respect to the higher-energy one, polarized along the long molecular axis, which is visible in the spectrum above 5~eV (see Fig.~\ref{fig.ant.el}a).

Moving now to the result obtained for the bimolecular cluster (Fig.~\ref{fig.ant.opt}b), we notice essentially the same spectral features discussed above for the isolated moiety, except that in this case, the lowest-energy peak is formed by two almost degenerate excitations.
Their energy separation, of the order of 10~meV in this system, corresponds to the Davydov splitting.
The electron and hole densities of the two excitations giving rise to the first peak [see vertical bars in Fig.~\ref{fig.ant.opt}b), left panel] are localized on the same molecule.
The lowest-energy excitation stems from the transition between the HOMO and the LUMO+1, while the second one from HOMO-1$\rightarrow$LUMO.
Their polarization is along the short molecular axis, as in the isolated molecule. 
Excitations polarized along the long molecular axis appear at higher energies, above 5~eV.

The absorption spectrum calculated for the crystal is plotted showing two diagonal components of the dielectric tensor (see Fig.~\ref{fig.ant.opt}c).
A sharp resonance along $b$ dominates the onset followed by a very weak excitation along $c$, in agreement with earlier results obtained at the same level of theory~\cite{humm+04prl}.
The presence of two excitations with two polarizations is due to Davydov splitting.
However, in the crystal, this effect is not only related to the presence of two inequivalent molecules in the simulation cell, as in the cluster model, but also to their mutual orientation with respect to the crystalline axes.
Appropriate rotation of the dielectric tensor along the electric field directions enhances the spectral intensity of the two Davydov components~\cite{cocc+18pccp,vorw+16cpc}.
The spatial distribution of the electron-hole pair corresponding to the lowest-energy exciton of the ANT crystal is shown as the square modulus of exciton wave function, see Eq.~\eqref{eq:exciton}. 
In Fig.~\ref{fig.ant.opt}c), the isosurface represents the correlated electron distribution with respect to the fixed hole position, marked in the plot by a red dot.
In analogy with its longer oligoacene siblings~\cite{humm-ambr05prb,cocc+18pccp}, lowest-energy exciton of the ANT crystal has Frenkel character.
Although this result may suggest that the adopting a cluster model is sufficient to capture the exciton nature in this system [compare Fig.~\ref{fig.ant.opt}b) and c)], the localization of the electron-hole pair in the crystal is actually a consequence of the large the electron-hole coupling strength, as extensively discussed in previous work on oligoacenes~\cite{humm-ambr05prb,cocc+18pccp,cuda+12prb}.
Neglecting the proper treatment of these effects, as provided by the solution of the BSE, would lead to an incorrect description not only of the optical spectra but especially of the exciton distribution, which would erroneously retain the periodicity of the electronic wave functions.

\section{Accounting for electrostatic interactions from an implicit environment}\label{sec:pcm}

In the previous section, we have shown the importance of describing an organic material in its actual state of matter in order to capture correctly its physical properties.
Long-range effects such as band dispersion and optical response require accounting explicitly for the lattice periodicity.
Modelling the system only through the two (isolated) molecules included in the unit cell renders solely those properties that are ruled by short-range (\textit{i.e.}, nearest-neighbor) interactions.
Yet, the full-fledged simulation of an organic crystal is not a trivial task. 
Common bottlenecks are the availability of input structures and of the computational resources that enable performing converged calculations on such systems with state-of-the-art \textit{ab initio} methods.
In the following, we see how implicit schemes, accounting only for electrostatic interactions between molecules and their isotropic as well as anisotropic environment and  applied in conjunction with DFT, can yield a correct estimate of the band gap of organic materials in different configurations.

\subsection{Molecules in isotropic environments: The best-case scenario for PCM}

\begin{figure}
    \centering
    \includegraphics[width=0.75\textwidth]{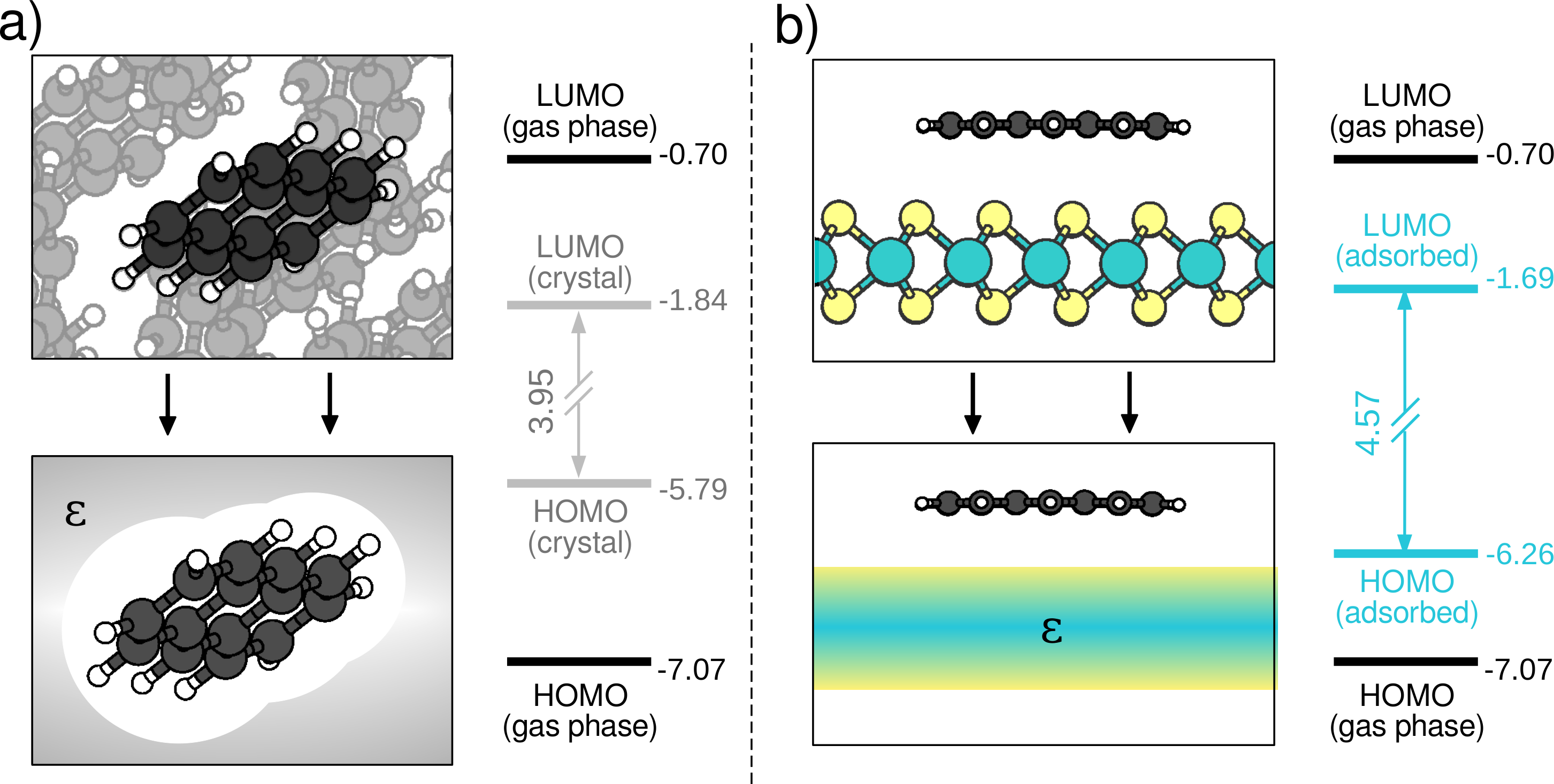}
    \caption{Replacing the environment by a polarizable continuum described by the static dielectric function $\varepsilon$: a) anthracene crystal simulated in the framework of conventional PCM~\cite{toma+05cr} assuming one molecule embedded in a homogeneous cavity, and b) anthracene molecule physisorbed on a MoS$_2$ monolayer modeled with \textsc{LayerPCM}~\cite{krum+21jcp}. The energies (in eV) of the frontier orbitals, HOMO and LUMO, are estimated as -IP and -EA, respectively. }
    \label{fig:ant-pcm}
\end{figure}

As a first example, we calculate the band-gap of the ANT crystal simulating atomistically only a single moiety and accounting implicitly for the surrounding molecules with the aid of the PCM.
In this analysis, instead of computing the fundamental gap as the difference between LUMO and HOMO energies, we assume the definition $E_{gap} = \text{IP} - \text{EA}$. 
IP and EA are the ionization potential and electron affinity of the molecule, respectively, calculated as the total energy difference between the neutral species and its cation (ANT$^+$) and anion (ANT$^-$): $\text{IP} = E_{tot}^{ANT^+} - E_{tot}^{ANT}$ and $\text{EA} = E_{tot}^{ANT} - E_{tot}^{ANT^-}$.
This corresponds to the so-called $\Delta$SCF approach, which we apply herein adopting the GGA for the exchange-correlation functional (see computational details in the Appendix).

We start from the gas-phase molecule, for which we obtain IP $=7.07$~eV and EA$= 0.70$~eV, both in good agreement with corresponding experimental data: measurements reported for IP range from 7.40 to 7.45~eV~\cite{schmidt1977jcp, boschi1974jcp}, while for EA, reference values are between 0.53 and 0.66~eV~\cite{heinis+1993oms, ruoff+1995jpc, naoto+2007jcp}. 
The fundamental gap of gas-phase ANT is therefore equal to 6.37~eV (see Fig.~\ref{fig:ant-pcm}a, right), \textit{i.e.}, slightly smaller than the experimental prediction (6.74 -- 6.92~eV) extracted from the values of EA and IP reported above and reported in Ref.~\cite{prob-karl75pssa}.
Discrepancies between calculated and measured values can be ascribed to the adopted GGA for the exchange-correlation functional. 

Next, we apply the same procedure embedding the ANT molecule, its anion, and its cation into a PCM cavity filled by a homogeneous medium with $\varepsilon=3.10$, corresponding to the static dielectric constant computed for the ANT crystal in the random-phase approximation (RPA). 
In this setup, the calculated IP decreases to 5.79~eV and the EA increases to 1.84~eV (see Fig.~\ref{fig:ant-pcm}a, right).
It is interesting to notice that both quantities vary by an almost identical amount in comparison to their values in gas phase. 
The fundamental gap becomes 3.95~eV, in excellent agreement with available measurements for transport gaps in ANT~\cite{baes+kill1973mclc,belk+grec1974pssa,riga+1977phsc}. 
It is worth noting that this good match is enabled by the low dispersion of the bands in the ANT crystal (see Fig.~\ref{fig.ant.el}c), which is in turn an consequence of the localized character of electronic states of the constituents.
In the presence of highly dispersive bands, where the energy levels assume very different values at varying \textbf{k}-vector, such a good agreement with an effective model for the environment is not foreseeable.
Band dispersion is the consequence of precisely the type of intermolecular interactions that are neglected in the PCM and that require explicit quantum-mechanical modelling to be captured.

\begin{table}
    \centering
    \caption{Dielectric constants ($\varepsilon$) computed from the RPA for the corresponding crystals, energy gaps (E$_{gap}$) of naphthalene (NAP), anthracene, (ANT), and tetracene, (TET) calculated with the $\Delta$SCF method applied to the isolated moieties in the neutral and charged ($\pm$1) states supplemented by PCM to mimic the crystalline environment, and experimental reference values: $^a$ Ref.~\cite{baes+kill1973mclc}; $^b$ Ref.~\cite{belk+grec1974pssa}; $^c$ Ref.~\cite{riga+1977phsc}. 
    }
    \vspace{0.5cm}
    \begin{tabular}{>{\centering\arraybackslash}p{2cm}|>{\centering\arraybackslash}p{2cm}|>{\centering\arraybackslash}p{3cm}|>{\centering\arraybackslash}p{3cm}}
        System & $\varepsilon$ & E$_{gap}^\mathrm{\Delta SCF}$ (eV) & E$_{gap}^\mathrm{exp.}$ (eV) \\
        \hline
        NAP &  2.80 & 5.30 & 5.40$-$5.53~$^{a,b}$  \\
        ANT & 3.10& 3.95 & 3.90$-$4.00~$^{a,b,c}$ \\
        TET & 3.29 & 3.02 & 3.00$-$3.11~$^{a,b,c}$ \end{tabular}
    \label{tab:acenes-pcm}
\end{table}

To prove that the results obtained for ANT are not just a favorable coincidence, we apply the same procedure to other short members of the oligoacene family. 
The results reported in Table~\ref{tab:acenes-pcm} and the shown agreement with experimental references confirm the ability of this approach to correctly reproduce the band gaps of the crystals.
It is worth noting that the static dielectric functions used to feed the PCM in these calculations have been computed from the RPA for the corresponding crystalline phases.
However, this is not the only way to retrieve these quantities.
A cheap alternative is to evaluate them from linear-response calculations on the gas-phase moiety and from the Clausius-Mossotti equation, which connects the microscopic polarizability of the single unit with the macroscopic dielectric constant of the crystal~\cite{hu+17jcc, zhu+18jpcc}.
Otherwise, for the considered systems, which are well-known and extensively characterized, corresponding empirical values are available in the literature~\cite{belk+grec1974pssa,pope-swen99book}.

The generally good accuracy of the adopted $\Delta$SCF approach in the prediction of IP and EA makes viable the usage of optimally tuned, Koopman-compliant hybrid functionals~\cite{tamar+2009jacs}. Parameters therein are adjusted in a non-empirical way until the HOMO energy matches the $\Delta$SCF-determined IP; the EA can be accounted for as well, demanding that it agree with either the LUMO energy of the neutral molecule or - more strictly compliant with Janak's theorem~\cite{jana1978prb} - the HOMO energy of the anion. Actually, the IP and EA obtained with $\Delta$SCF are relatively robust with respect to the replacement of DFT with Fock exchange, meaning that hybrid functionals do not necessarily improve very much the estimate of these values with respect to pure DFT. However, hybrid functionals - particularly the long-range-corrected ones with a large amount of Fock exchange in the asymptotic limit - are essential when computing optical excitations with TDDFT, enabling acceptably accurate results even for charge-transfer excitation energies, which pure DFT notoriously struggles with~\cite{zhen+17jpcl}. Such hybrid TDDFT approaches coupled with the PCM can yield reasonable estimates for bandgaps and even for exciton binding energies~\cite{hu+17jcc,zhu+18jpcc}, in spite of neglecting static intermolecular interactions, and artificially confining excitons, \textit{e.g.}, into a single unit cell. 
These approximations can be softened by replacing a single molecule by a cluster, thus describing a larger portion of the system atomistically and quantum-mechanically. In this case, some pitfalls with respect to symmetry have to be avoided, which is generally lower for clusters than for both isolated molecules and the organic crystals~\cite{craciunescu+2022jpcl}.

\subsection{Repurposing the PCM: From solvent cavities to substrate layers}

Using the generic prescription of Eq.~\eqref{eq:pcm_mat} for building the PCM matrix, one can model even more complex dielectric environments than an isotropic solvation cavity. The only requirement is an expression for the screened Coulomb interaction $W$. In Ref.~\cite{krum+21jcp}, we extended the PCM to describe implicitly an arbitrary stack of layered materials intrinsically characterized by a high degree of anisotropy, introducing \textsc{LayerPCM}. This scheme reproduces the increasingly popular scenario of organic molecules adsorbed on two-dimensional transition-metal dichalcogenides (TMDC)~\cite{jari+16nl,liu+17nl,song+17nano,zhan+18am,gu+18acsp,amst+19nano,park+21as}. In this case, the screened interaction in the region above the slab can be expressed as 
\begin{align}
    W(\textbf{r},\textbf{r}')= v_c(\textbf{r}-\textbf{r}') + W^p(\textbf{r},\textbf{r}'),
\end{align}
where the so-called polarization contribution, $W^p$, is related to the polarization of the substrate. For a semi-infinite substrate, $W^p$ can be written as the electrostatic potential of a single image charge within the substrate region. For a stacked substrate, it is more convenient to calculate it numerically as
\begin{align}\label{eq:W_LayerPCM}
    W^p(\textbf{r},\textbf{r}') = \int_0^\infty\text dq_\parallel\, J_0(q_\parallel\Delta r_\parallel)\frac{1-\varepsilon(q_\parallel)}{1+\varepsilon(q_\parallel)}e^{-q_\parallel(z'+z-2z_0)},
\end{align}
using a Gauss-Laguerre quadrature~\cite{abra1974}. In Eq.~\eqref{eq:W_LayerPCM},
$\Delta r_\parallel = [(x-x')^2+(y-y')^2]^{1/2}$, $J_0$ is the zeroth-order Bessel function of first kind \cite{temme1996}, and $\varepsilon(q_\parallel)$ is the effective nonlocal dielectric constant of the substrate, which can be determined using a transfer matrix formalism \cite{krum+21jcp}. An alternative to this Bessel-function-based approach for obtaining $W^p$ is the recursive calculation of image charges \cite{kuma-taka1989prb}.

We apply \textsc{LayerPCM} to model the hybrid interface formed by an ANT molecule adsorbed on a MoS$_2$ monolayer (see Fig.~\ref{fig:ant-pcm}b).
Three parameters are required to model the TMDC (or any other layered material) within \textsc{LayerPCM}: the layer thickness $t$, as well as the perpendicular and parallel components of the static dielectric constant, $\varepsilon_\perp$ and $\varepsilon_\parallel$, respectively. We employ a self-consistent scheme to obtain these quantities from first principles. The MoS$_2$ monolayer is simulated atomistically with periodic boundary conditions in all three directions, inserting a large amount of vacuum in the perpendicular direction to decouple the replicas. The dielectric constants are determined through linear-response calculations within the RPA. It should be noted that corresponding results are effective values, $\varepsilon_{\perp,\mathrm{eff}}$ and $\varepsilon_{\parallel,\mathrm{eff}}$, describing simultaneously the TMDC and the vacuum layer. In order to extract the dielectric constants of the material excluding the vacuum, we take an initial guess for $t$ and perform the following transformations:
\begin{subequations}\label{eq:eps_3d_to_2d}
\begin{align}
    \varepsilon_\parallel &= \left(\varepsilon_\parallel^\mathrm{eff}-1\right)\frac{c}{t}+1\\
    \varepsilon_\perp &= \left[\left( \varepsilon_{\perp,\mathrm{eff}}^{-1}-1\right)\frac{c}{t}+1\right]^{-1}, \label{eq:eps_3d_to_2d_perp}
\end{align}
\end{subequations}
where $c$ is the perpendicular lattice constant of the adopted supercell, \textit{i.e.} the sum of the thickness of the TMDC and the height of the vacuum. The different formulas adopted for the in-plane and out-of-plane directions is due to the different boundary conditions for electric fields with parallel and perpendicular polarization with respect to the dielectric interfaces~\cite{apsn1982ajp}. Equipped with these values, $t$ is obtained by connecting smoothly the plane-averaged semi-local $v_{xc}$ at short range with the image potential $V_\mathrm{im}$ of the MoS$_2$ layer at long range, corresponding to the correct asymptotic tail of the exact exchange-correlation potential~\cite{egui-hank1989prb}. This approach was employed in the past to determine mirror planes for metal surfaces~\cite{lang-kohn1973prb}. For the dielectric monolayer, the image potential follows as a special case of Eq.~\eqref{eq:W_LayerPCM} with $\textbf{r}=\textbf{r}'$ and $\varepsilon(q_\parallel)$ corresponding to a single layer:
\begin{align}\label{interactionThickness.eq}
   V_\mathrm{im}(z) &= -\frac 12[(\varepsilon_\perp\varepsilon_\parallel)^{1/2}+1][(\varepsilon_\perp\varepsilon_\parallel)^{1/2}-1]\times\nonumber\\&\times\int_0^\infty\text dq_\parallel\,\frac{e^{-2q_\parallel z}}{1+\varepsilon_\perp\varepsilon_\parallel+2(\varepsilon_\perp\varepsilon_\parallel)^{1/2}\coth(q_\parallel(\varepsilon_\parallel/\varepsilon_\perp)^{1/2} t)}.
\end{align}
The value obtained for $t$ by smoothly matching $v_{xc}$ and $V_\mathrm{im}$ is reinserted into Eq.~\eqref{eq:eps_3d_to_2d} to calculate new dielectric constants. This procedure is iterated self-consistently. For MoS$_2$, we obtain $\varepsilon_\parallel=16.64$, $\varepsilon_\perp=11.25$, and $t=5.4~\mathrm{\AA}$. Owing to the prescription in Eq.~\eqref{eq:eps_3d_to_2d_perp}, $\varepsilon_\perp$ is extremely sensitive to changes in $t$. Indeed, values for $\varepsilon_\perp$ differing by a factor of 2 have been reported in Ref.~\cite{latu+2018npj} using a different definition of $t$. While such a wide range of results may seem alarming, it can be understood by realizing that the dielectric constant does not single-handedly describe the polarization response of the medium, but the geometry plays a crucial role, too. This is particularly true when the dimensions of the medium are small. Very different combinations of $\varepsilon_\perp$ and $t$ can thus give rise to similar physical scenarios.

Having defined this scheme, we evaluate the band gap of the physisorbed ANT molecule, where the underlying layer is accounted for via its dielectric function only.
As in the crystalline environment, the presence of the TMDC reduces significantly the band gap of the moiety. This is a well-known effect that has been reported in a number of first-principles studies on hybrid interfaces formed by conjugated molecules adsorbed on TMDC monolayers~\cite{chou+17jpcc,habi+20ats,aden2021jcp,mela+22pccp,oliv+22prm,guo+22nr}. 
Comparing the results obtained for isolated ANT and for ANT adsorbed on a MoS$_2$ monolayer, we notice a reduction of IP and an increase of EA by about 0.8 and 1~eV respectively, which lead to a fundamental gap of 4.57~eV for the physisorbed molecule (see Fig.~\ref{fig:ant-pcm}b, right panel).
It is worth noting that the difference between IP and EA implies that already the relatively small ANT ions cannot be assimilated to point charges, for which both values would be equal.
The band-gap renormalization for the molecule adsorbed on the TMDC is lower than in the crystal. Comparing the results reported in Fig.~\ref{fig:ant-pcm}, panels a) and b), this behavior can be ascribed to more significant energetic readjustement of the HOMO level (set equal to -IP, according to Janak's theorem) in the two scenarios.
In contrast, the LUMO level, estimated as -EA, differs energetically by 150~meV in the implicit simulation of the crystalline environment and of the monolayer substrate.

\begin{figure}[h!]
    \centering
    \includegraphics[width=0.95\textwidth]{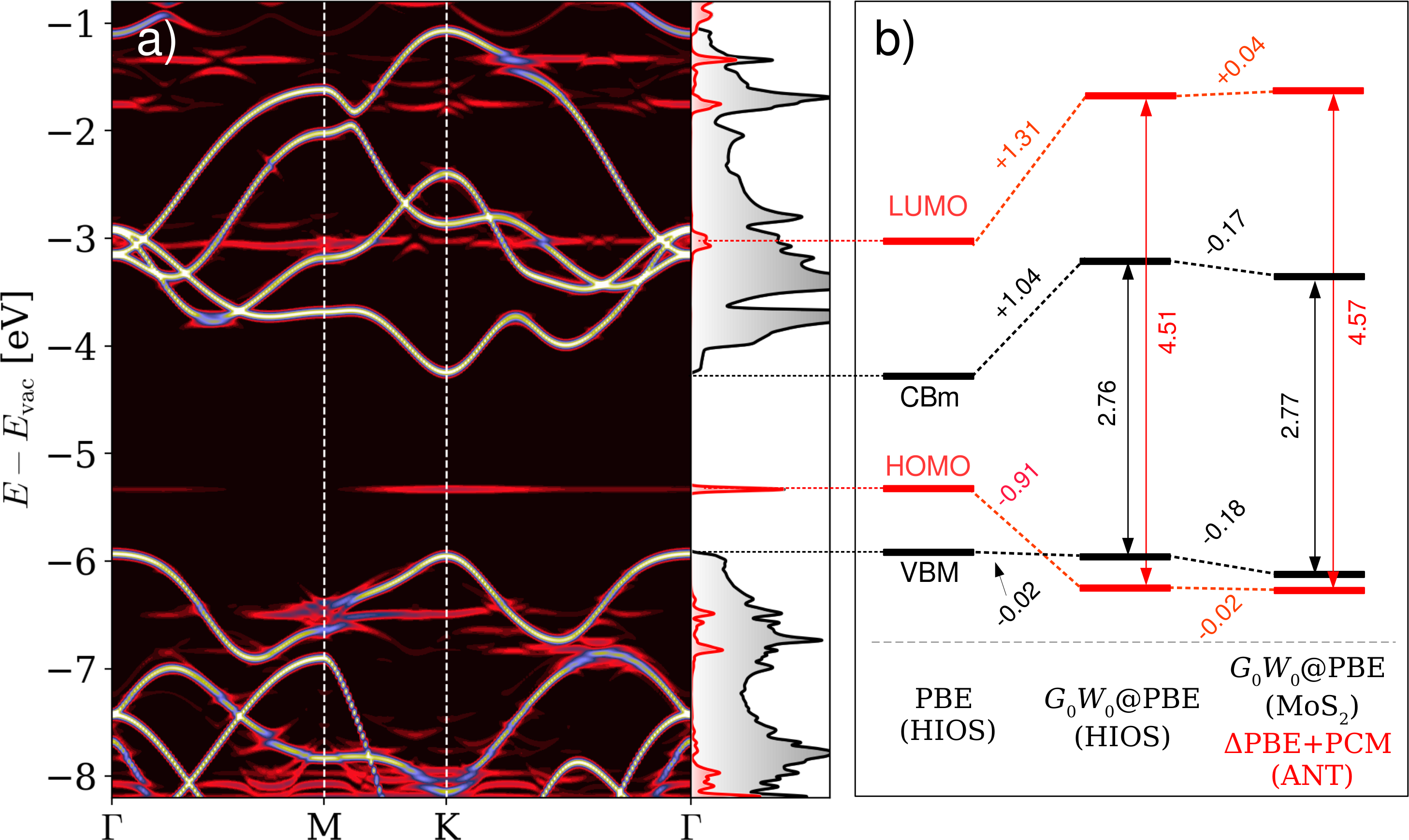}
    \caption{Electronic structure of ANT adsorbed on a MoS$_2$ monolayer. a) Band structure and density of states (DOS) computed from DFT (PBE functional) for the atomistically modeled hybrid interface. The band structure of the composite system is unfolded in the unit cell of the free-standing monolayer, whose electronic states are shown in gold. The spectral weights of the energy levels of the hybrid interface range from red (low values, indicating contributions of molecular states) to indigo (high values, associated with hybridized states). In the DOS, contributions from the molecular (all) states are shown in red (black), b) Level alignment of the frontier levels of ANT (red) and MoS$_2$ (black) computed from different levels of theory: PBE and $G_0W_0$-corrected PBE for the complete hybrid system, plus the combination of two subsystem calculations, using $G_0W_0@$PBE for MoS$_2$ and $\Delta$PBE ($\Delta$SCF with PBE functional) with PCM for ANT.}
    \label{fig:ant-mos2}
\end{figure}

To analyze more in depth the electronic properties of ANT physisorbed on the MoS$_2$ monolayer, we perform a fully atomistic benchmark calculation of the whole hybrid system simulating a supercell corresponding to 4$\times$4 the unit cell of the TMDC and with the molecule adsorbed flat at an approximate distance of 3.3~\AA{} from the substrate. 
The electronic structure of the system, evaluated for consistency with the previous runs at the GGA level of DFT (PBE functional, see computational details in the Appendix), features a type-II band alignment between the organic and inorganic components (see Fig.~\ref{fig:ant-mos2}). The HOMO of the molecule is situated within the energy gap of MoS$_2$, while the LUMO is higher up in the conduction band. We note in passing the clear signatures of mixing between the HOMO-1 of ANT and the wave functions of MoS$_2$ at the high-symmetry point $M$ in the valence region, which was found to be a hybridization hotspot between TMDCs and polycyclic aromatic hydrocarbons~\cite{krum-cocc21es}. 
It is well known that DFT - particularly with semi-local functionals - is not very good at predicting energy level alignments: it notoriously underestimates band gaps as well as molecular IPs and, furthermore, does not capture image-charge induced level renormalization~\cite{neaton+2006prl, garc+2009prb}. Thus, we correct the frontier energy levels of the hybrid system as well as of the freestanding monolayer for reference with many-body perturbation theory, using the single-shot $G_0W_0$ method (Fig.~\ref{fig:ant-mos2}b). As expected, this leads to a band-gap opening; the fundamental gap of MoS$_2$ is increased from 1.70 to 2.76~eV in the hybrid system, and to 2.77~eV in the freestanding configuration, in agreement with previously reported results on a similar level of theory~\cite{sokl+2014apl, huse+2013prb, qin+20jmcc, ryou+2016sr}. The presence of the molecule does not lead to a significant band-gap renormalization in the TMDC. The most important change brought about by the QP correction is the inverted order of the occupied energy levels. The severe underestimation of the ionization energy of ANT induced by DFT in the GGA is corrected by a large downshift of the HOMO, while the valence-band maximum of MoS$_2$ barely changes, resulting in a change from a type-II to type-I alignment with the molecular levels hugging those of the TMDC. Similar observations have been made for other hybrid interfaces \cite{aden2021jcp,oliv+22prm}. Hence, DFT alone yields unreliable results for the alignment of occupied states; for the unoccupied ones, however, the situation is not as dramatic, considering that the $GW$ corrections to the conduction-band minimum and to the LUMO differ only by 0.27~eV, compared to 0.89~eV between the valence-band maximum and the HOMO.

Interestingly, using the $\Delta$SCF+\textsc{LayerPCM} approach described above for calculating the renormalized band gap of ANT, we obtain results in excellent agreement with the $GW$-corrected ones, even adopting a GGA functional for DFT (Fig.~\ref{fig:ant-mos2}b). Taking as a point of comparison the frontier levels of isolated MoS$_2$ computed from $G_0W_0$, $\Delta$SCF+\textsc{LayerPCM} yields a significantly better estimate for the level alignment of the hybrid interface than DFT alone. The only salient difference between the fully atomistic $GW$ calculation on the hybrid system and the effective treatment is given by a rigid shift of the MoS$_2$ levels, which is brought about by charge transfer~\cite{mela+22pccp}.
Since the Fermi energy of ANT is higher than that of MoS$_2$, some electronic population is transferred from the organic to the inorganic side of the interface. The resulting partial charges - positive on ANT, negative on MoS$_2$ - slightly realign the levels of the subsystems with their corresponding electrostatic energy. This effect is obviously not captured when describing the two subsystems individually. However, the amount of charge transferred is based on the level alignment obtained from DFT in the GGA, as we apply only the perturbative $G_0W_0$ approach and not the self-consistent flavor of $GW$. The staggered line-up predicted by GGA can be assumed to lead to an overestimation of the difference between the individual Fermi energies and, thus, of charge transfer. Ongoing work to perturbatively include charge-transfer effects based on the energy levels predicted by calculations on the individual the subsystems is expected to further improve the accuracy of this method.

\section{Donor/acceptor co-crystals}\label{sec:co-cry}

In the last part of this review, we focus on another type of organic materials, namely donor/acceptor complexes and co-crystals.
In contrast with monomolecular compounds like ANT, in these systems, a complex interplay between short- and long-range interactions rules the electronic and optical responses~\cite{guer+21jpcc,vale+20pccp}.
In this analysis, we consider the prototypical system formed by quaterthiophene (4T) doped by the electron acceptor tetracianoquinodimethane (TCNQ).

The synthesis and characterization of doped organic films has been the subject of intensive research in the last decade~\cite{mend+15ncom,ping+10jpcl,mend+13acie,goet+14jpcc,kief+19natm,theu+21jpcc,russ+22jpcc}.
Depending on the chemical nature of the constituents and their mutual interactions, two main doping mechanisms have been identified, namely \textit{integer} and \textit{partial} charge transfer~\cite{salz+16acr}.
In the former case, an entire charge carrier is transferred from the donor to the acceptor, leading to efficient (opto)electronic performances.
In the latter scenario, the frontier orbital hybridization gives rise to a strong coupling between the components enabling a fractional charge transfer between them.
4T:TCNQ belongs to the second class of systems, also known as charge-transfer complexes (CTC).
The synthesis of CTC as ordered co-crystals is still in its infancy. One of the first, successful attempts~\cite{sato+19jmcc} enabled the resolution of the crystal structure of 4T doped by TCNQ with 1:1 ratio, which is considered in the following. 


\subsection{Electronic properties: Local interfaces vs. long-range effects}\label{sec:d/a}

\begin{figure}[h]
    \centering
\includegraphics[width=0.75\textwidth]{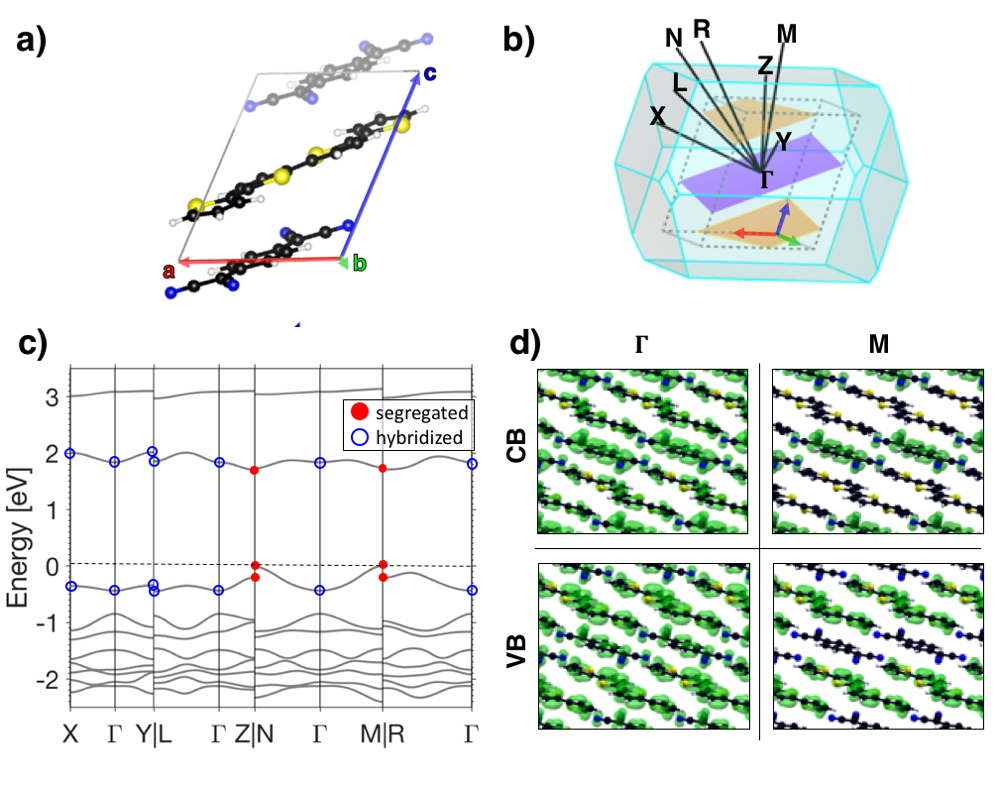}
    \caption{Top: a) Unit cell of the 4T:TCNQ co-crystal: carbon atoms are dark grey, sulphur atoms yellow, nitrogen atoms blue, and hydrogen atoms white. b) BZ of the co-crystal with the directions connecting $\Gamma$ to the other high-symmetry points marked by black solid lines. The unit cell of the co-crystal is represented inside the BZ by gray dotted lines including the molecular planes of the donor (violet) and acceptor molecules (gold). c) Band structure of the 4T:TCNQ co-crystal. d) Probability density of the highest valence (VB) and lowest conduction (CB) states at $\Gamma$ and $M$, where hybridization and segregation are visible, respectively.} 
\label{fig:4TTCNQ}
\end{figure}

4T:TCNQ crystallizes in a triclinic lattice with unit cell containing a donor/acceptor pair with basal planes facing each other~\cite{sato+19jmcc} (see Fig.~\ref{fig:4TTCNQ}a).
The complex arrangement of the molecules with respect to the crystal structure and their relation with the BZ is schematically visualized in Fig.~\ref{fig:4TTCNQ}b), where the directions connecting $\Gamma$ with the other high-symmetry points are highlighted to ease the reading of the band-structure plot (Fig.~\ref{fig:4TTCNQ}c).
By inspecting this result, we immediately notice highest-occupied and lowest-unoccupied states sticking out with respect to the manifold of valence and conduction bands, respectively.
Interestingly, the character of these states varies at different \textbf{k}-points in the BZ.
At the zone edges (high-symmetry points $Z$, $N$, $M$, and $R$), the wave functions are spatially segregated over the donor (valence states) and acceptor molecules (conduction states), respectively, see Fig.~\ref{fig:4TTCNQ}d).
Notably, both valence-band maximum and conduction-band minimum are located at $M$. 
Elsewhere in the BZ, the electronic states are hybridized between the donor and the acceptor, in analogy with the HOMO and the LUMO of the corresponding molecular complexes~\cite{zhu+11cm,mend+15ncom}.

\begin{figure}
    \centering
    \includegraphics[width=0.8\textwidth]{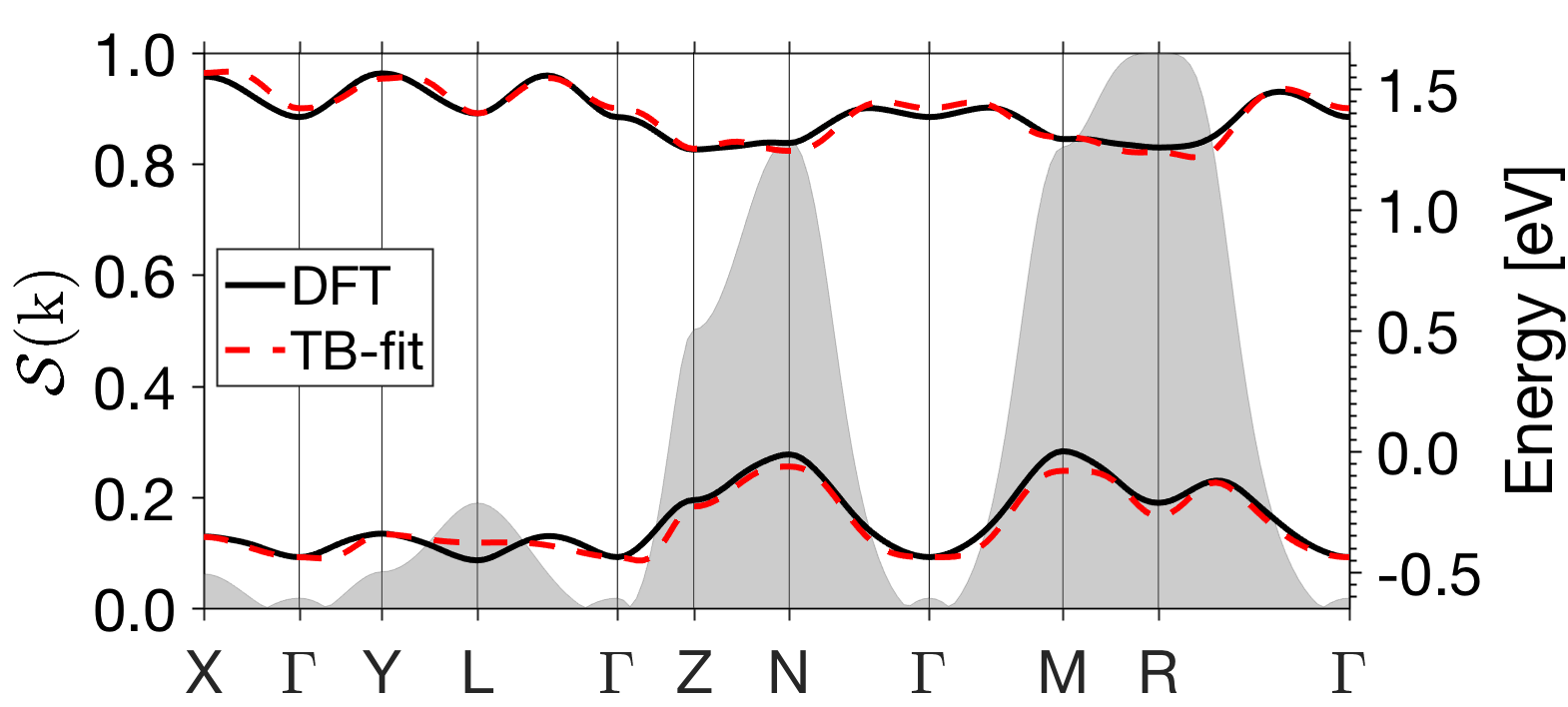}
    \caption{Highest valence band and lowest conduction band of 4T:TCNQ computed from DFT (black solid line) and from the TB fit (red dashed line); the corresponding segregation factor $\mathcal{S}(\mathbf{k})$, see Eq.~\eqref{eq:sf}, is given to the gray shaded area.}
    \label{fig.tb}
\end{figure}

This counter-intuitive behavior of the frontier states in the co-crystal can be explained by the long-range nature of its electronic wave functions~\cite{guer+21jpcc}. 
To this end, we introduce a tight-binding (TB) model assuming as a basis the molecular orbitals of the isolated bimolecular donor/acceptor cluster extracted from the crystalline unit cell.
We assimilate the system to a quantum-mechanical two-level model including the highest-occupied and lowest-unoccupied energy levels, with corresponding wave functions expressed as
\begin{equation}
    \label{tb-bloch}
    \psi_{\lambda\textbf{k}}(\textbf{r}) = \sum_{m=H,L} C^{\lambda}_m(\textbf{k})\tilde{\varphi}_{m\textbf{k}}(\textbf{r}),
\end{equation}
where $C^{\lambda}_m(\textbf{k})$ are the orthonormal expansion coefficients for either level $\lambda$, and
\begin{equation}
    \label{tb-bloch2}
    \tilde{\varphi}_{m\textbf{k}}(\textbf{r}) = \sum_j \varphi_m(\textbf{r} + \textbf{R}_j) e^{-i\textbf{k}\cdot \textbf{R}_j}.
\end{equation}
In Eq.~\eqref{tb-bloch2}, $\varphi_m(\textbf{r})$ are the HOMO ($m=H$) and LUMO ($m=L$) of the isolated donor/acceptor complex, and $\textbf{R}_j$ are the lattice vectors. 
From Eqs.~\eqref{tb-bloch} and \eqref{tb-bloch2}, it is evident that $\psi_{\lambda\textbf{k}}(\textbf{r})$ has the same periodicity of the co-crystal and thus fulfills Bloch's theorem.
It is worth highlighting the \textbf{k}-point dependence of the expansion coefficients which brings about the \textbf{k}-dependent character of the electronic states within each band seen in Fig.~\ref{fig:4TTCNQ}c).
The wave functions and energies associated to the valence and conduction bands are obtained by diagonalizing the following problem:
\begin{equation}
\hat{H}^{TB} \begin{pmatrix} C_H^{\lambda}(\textbf{k}) \\ C_L^{\lambda}(\textbf{k}) \end{pmatrix} =
\begin{pmatrix} A_{HH}(\textbf{k}) & A_{HL}(\textbf{k}) \\ A_{HL}(\textbf{k}) & A_{LL}(\textbf{k})  \end{pmatrix}
\begin{pmatrix} C_H^{\lambda}(\textbf{k}) \\ C_L^{\lambda}(\textbf{k})  \end{pmatrix}
=E_{\lambda}(\textbf{k})
\begin{pmatrix} C_H^{\lambda}(\textbf{k}) \\ C_L^{\lambda}(\textbf{k}) \end{pmatrix},
\label{tbeq}
\end{equation}
where the nearest-neighbor (NN) approximation is assumed and the overlap integral is $\sum_j e^{-i\textbf{k} \cdot \textbf{R}_j}\int \varphi_n(\textbf{r})\varphi_m(\textbf{r}+\textbf{R}_j)d^3r \approx \delta_{mn}$. 
Both assumptions are justified by the overall localized nature of the electronic states in molecular crystals. 
The matrix elements on the left-hand-side of Eq.~\eqref{tbeq} are calculated as
\begin{equation} \label{Anm}
    A_{nm}(\textbf{k}) = A_{mn}(\textbf{k}) = \sum_{j \in NN} e^{-i\textbf{k} \cdot \textbf{R}_j}\int\text d^3r\, \varphi_n^*(\textbf{r}) \hat{H}^{TB} \varphi_m(\textbf{r} + \textbf{R}_j)  = t^{(0)}_{nm} + 2\sum_{j \in NN} t_{nm}^{(j)} \cos(\textbf{k} \cdot \textbf{R}_j),
\end{equation} 
where $t_{nm}^{(j)} = \int\text d^3r\,\varphi_n^*(\textbf{r}) \hat{H}^{TB} \varphi_m(\textbf{r} \pm \textbf{R}_j)  = \int\text d^3r\, \varphi_n^*(\textbf{r} \pm \textbf{R}_j) \hat{H}^{TB} \varphi_m(\textbf{r})$ for $n,m\in\{H,L\}$
are the on-site ($\textbf{R}_j=0$) and hopping ($\textbf{R}_j = \textbf{R}_a, \textbf{R}_b, \textbf{R}_c$) integrals. 
The eigenvectors of Eq.~\eqref{tb-bloch} for the wave functions of the valence (VB) and conduction bands (CB) are expressed as
\begin{equation} \label{eq:Cvb}
 C^{VB}_H(\textbf{k})=w(\textbf{k}),\;\;\;\;\; C^{VB}_L(\textbf{k})= -w(\textbf{k})\mu(\textbf{k})   
\end{equation} 
and  
\begin{equation} \label{eq:Ccb}
 C^{CB}_H(\textbf{k})=w(\textbf{k})\mu(\textbf{k}),\;\;\;\;\; C^{CB}_L(\textbf{k})=w(\textbf{k}), 
\end{equation}
respectively, where 
\begin{equation} \label{eq:mu}
\mu(\textbf{k}) = \frac{A_{HL}(\textbf{k})}{E_{(-)}(\textbf{k}) + \sqrt{E^2_{(-)}(\textbf{k}) + A^2_{HL}(\textbf{k})}}, 
\end{equation} 
with
\begin{equation} \label{eq:E-}
E_{(-)}(\textbf{k}) = A_{LL}(\textbf{k}) - A_{HH}(\textbf{k})
\end{equation} 
and
\begin{equation} \label{eq:W}
w(\textbf{k}) = \frac{1}{\sqrt{1 + \mu^2(\textbf{k})}}.
\end{equation} 
Plugging Eqs.~\eqref{eq:Cvb} and \eqref{eq:Ccb} into Eq.\eqref{tb-bloch}, we get:
\begin{equation} \label{eq:vbm}
    \psi_{VB,\textbf{k}}(\textbf{r}) = \sum_j e^{-i\textbf{k} \cdot \textbf{R}_j}w(\textbf{k}) \left[ \varphi_H(\textbf{r} + \textbf{R}_j) - \mu(\textbf{k})\varphi_L(\textbf{r} + \textbf{R}_j) \right],
\end{equation}
and
\begin{equation} \label{eq:cbm}
    \psi_{CB,\textbf{k}}(\textbf{r}) = \sum_j e^{-i\textbf{k} \cdot \textbf{R}_j}w(\textbf{k}) \left[ \mu(\textbf{k})\varphi_H(\textbf{r} + \textbf{R}_j) + \varphi_L(\textbf{r} + \textbf{R}_j) \right].
\end{equation}
The energies associated to these states are calculated from Eq.~\eqref{tbeq} as
\begin{equation} \label{eq:Ek}
    E_{\lambda}(\textbf{k})= \sum_{j }\sum_{n,m} C^{\lambda}_n(\textbf{k}) C^{\lambda}_m(\textbf{k}) t^{(j)}_{nm} \cos(\textbf{k} \cdot \textbf{R}_j)   \;\;\;\; (\lambda = VB, CB). 
\end{equation}
With this formula, we can finally plot the energies of the VB and CB fit from the TB model (Eq.~\ref{eq:Ek}) with those output by the DFT calculations on the co-crystal, see Fig.~\ref{fig.tb}, where we adopt a different \textbf{k}-path with respect to Fig.~\ref{fig:4TTCNQ}c) to ease visualization.
The agreement between the two results is very good and confirms the validity of the model, which, therefore, can be used to rationalize the varying spatial distribution of the VB and CB wave functions in the BZ.
To this end, we introduce the so-called \textit{segregation factor},
\begin{equation}
    \mathcal{S}(\mathbf{k})= 1 - w^2(\textbf{k})\left( 1 - |\mu(\textbf{k})| \right)^2 = \frac{2\mu(\textbf{k})}{1 + \mu^2(\textbf{k})} \in [0,1],
    \label{eq:sf}
\end{equation}
and we plot it in Fig.~\ref{fig.tb} for a direct visualization of the band character.
Large values of $\mathcal{S}(\mathbf{k})$, found in the vicinity of the high-symmetry points $M$, $R$ as well as $Z$ and $N$, correspond to spatial segregation of the wave function on either molecule.
On the other hand, the small magnitude of $\mathcal{S}(\mathbf{k})$ close to $X$, $\Gamma$, $Y$, and $L$ is indicative of orbital delocalization across the donor/acceptor interface. 

We can rationalize these results going back to the equations introduced above.
From Eq.~\eqref{eq:mu}, it appears that $\mu(\mathbf{k})$ is maximized when $A_{HL}(\textbf{k})$ is large, \textit{i.e.}, when there is non-negligible mixing between the HOMO and the LUMO in the donor/acceptor complex. This happens along the crystallographic orientations that are substantially collinear to the lattice vectors aligned with the stacking direction of the molecules (see Fig.~\ref{fig:4TTCNQ}b).
Since the off-diagonal elements of $\hat{H}^{TB}$ are non-negligible, the expansion coefficients of the wave-functions in Eqs.~\eqref{eq:Cvb}-\eqref{eq:Ccb} have similar magnitudes.
In this scenario, long-range interactions prevail over local couplings between donor and acceptor moieties, and induce the spatial segregation of the frontier states seen in Fig.~\ref{fig:4TTCNQ}d). 
On the other hand, along the directions that are approximately parallel to the molecular planes and thus orthogonal to the stacking direction, $A_{HL}(\textbf{k}) \rightarrow 0$ and no mixing between HOMO and LUMO takes place.
In this case, $\hat{H}^{TB}$ becomes diagonal, and one $C^{\lambda}_m(\textbf{k})$ coefficient dominates over the others.
Consequently, $\mu(\mathbf{k}) \rightarrow 0$ implying $\mathcal{S}(\mathbf{k}) \rightarrow 0$: local couplings prevail over the long-range interactions, and the hybridized character of the frontier states is preserved in the periodic wave functions (Fig.~\ref{fig:4TTCNQ}d).


\subsection{Optical Properties: Charge-transfer excitations enhanced}

\begin{figure}
    \centering
 \includegraphics[width=0.75\textwidth]{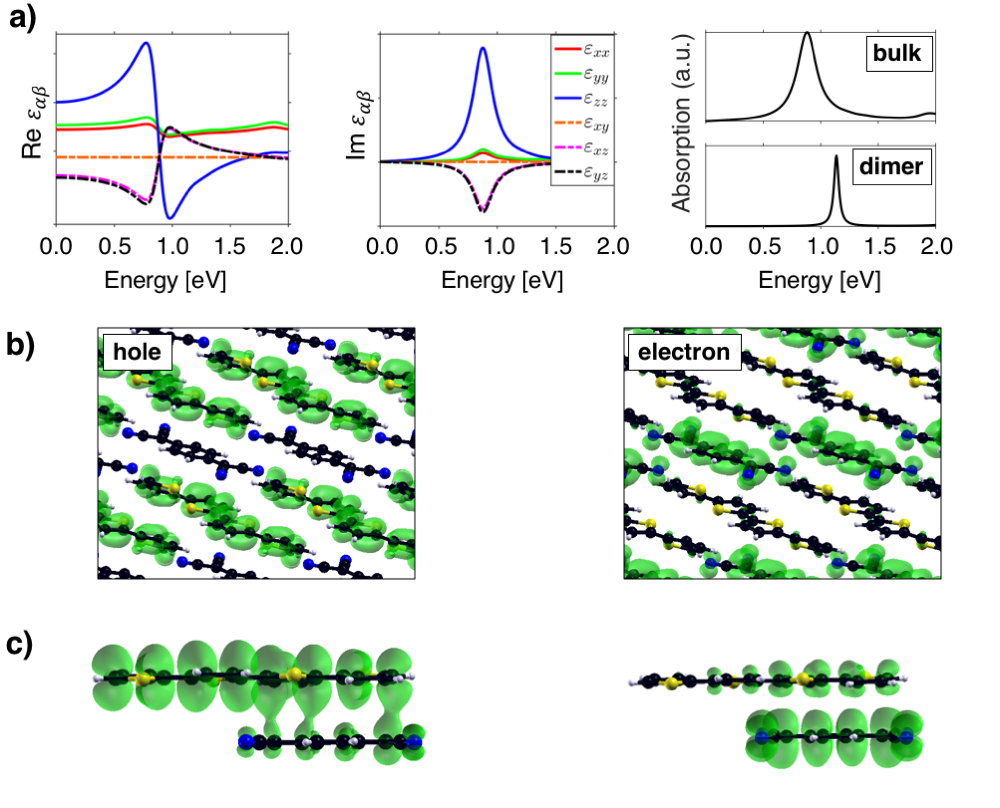}
    \caption{a) Real (left) and imaginary part (middle) of the full dielectric tensor of the 4T:TCNQ co-crystal computed from the BSE, including a Lorentzian broadening of 200~meV. On the right panels, the absorption spectra of the co-crystal [taken as the trace of the tensor shown in panel b)] and of the bimolecular cluster (dimer) extracted from it are shown; for the latter, a Lorentzian broadening of 100~meV is adopted. Hole (left panel) and electron (right panel) density of the first optical excitation b) in the co-crystal and c) in the cluster. 
    }
    \label{fig:epsM}
\end{figure}

With the gained understanding of the electronic properties of the 4T:TCNQ co-crystal, we now analyze its optical excitations.
Similar to ANT (Section~\ref{sec:ant} and Ref.~\cite{humm+04prl}) and to organic crystals in general~\cite{ruin+02prl,tiag+03prb,cocc+18pccp,cocc-drax15prb,cocc+16jcp,cocc-drax17jpcm,humm-drax05prb2}, the absorption spectrum of 4T:TCNQ is dominated by a strong resonance at the onset (see Fig.~\ref{fig:epsM}), on which we focus in the following.
The pronounced anisotropy of the co-crystal is apparent in both the real and imaginary part of the dielectric tensor computed from the BSE [Fig.~\ref{fig:epsM}a), left and middle panels]. 
All the six inequivalent components contribute to the lowest-energy excitation.
The one with the largest spectral strength is along $zz$, which is roughly aligned to the stacking direction of the molecules in the unit cell [see Fig.~\ref{fig:4TTCNQ}a)-b)].
This behavior is consistent with the first optical excitation of the isolated donor/acceptor complex~\cite{guer+21jpcc}, visualized in the spectrum in Fig.~\ref{fig:epsM}a), right panel.
The main contribution to the lowest-energy excitation in the co-crystal comes from the transition between the VB maximum and the CB minimum, again in line with the behavior of the cluster where the first excited state corresponds to the HOMO$\rightarrow$LUMO transition~\cite{guer+21jpcc}.
However, as discussed above, the wave functions of the frontier states are segregated on the donor and acceptor molecules, respectively (see Fig.~\ref{fig:4TTCNQ}d).
As a result, the first exciton in the co-crystal has charge-transfer character, as indicated in Fig.~\ref{fig:epsM}b) by the plots of the hole and electron densities, opposite to the Frenkel nature of its counterpart in the isolated complex (see Fig.~\ref{fig:epsM}c).
In light of this behavior, the strong intensity of first excitation of the co-crystal may appear puzzling.
To solve this conundrum, we make use again of the TB model introduced in Section~\ref{sec:d/a}.

Starting from the definition of Bloch's states (Eq.~\ref{tb-bloch2}), we peel off the periodic part
\begin{equation} \label{tdm3}
    u_{\lambda}(\textbf{k},\textbf{r})= \sum_j \sum_{m=H,L} C^{\lambda}_m(\textbf{k})\varphi_m(\textbf{r} + \textbf{R}_j) e^{-i\textbf{k}\cdot (\textbf{r} + \textbf{R}_j)}
\end{equation}
and use it to express the matrix elements for the optical transitions as
\begin{equation} \label{tdm1}
    \langle u_{CB}(\textbf{k},\textbf{r})|\hat{i\nabla_{\textbf{k}}} | u_{VB}(\textbf{k},\textbf{r}) \rangle = \sum_{m=H,L} C_m^{CB}(\textbf{k}) (i\nabla_{\textbf{k}}) C_m^{VB} ({\textbf{k}}) + \sum_{m,m'=H,L} C_{m'}^{CB}({\textbf{k}})C_m^{VB}({\textbf{k}}) \textbf{d}_{m'm},
\end{equation}
where 
\begin{equation} \label{tdm2}
    \textbf{d}_{m'm} = \int\text d^3r\, \varphi_{m'}^* (\textbf{r}) \, \hat{\textbf{r}} \, \varphi_m (\textbf{r})
\end{equation}
is the transition dipole moment between the (localized) molecular orbitals of the donor/acceptor cluster extracted from the unit cell of the co-crystal.
This term is not the only one entering the expression of the matrix elements.
The first term on the right-hand side of Eq.~\eqref{tdm1}, including the \textbf{k}-derivative of the expansion coefficients introduced in Eqs.~\eqref{eq:Cvb}-\eqref{eq:Ccb}, enables long-range couplings between the periodic states even when their spatial overlap is small. 
Thanks to this contribution, charge-transfer excitations like the lowest-energy one in the spectrum of the 4T:TCNQ co-crystal and of its siblings with partially or fully fluorinated acceptors~\cite{guer+21jpcc} exhibit finite oscillator strength. 
Notice that, in contrast, charge-transfer excitations in non-periodic donor/acceptor organic interfaces are characterized by very weak spectral intensity~\cite{theu+21jpcc,russ+22jpcc}, due to the negligible spatial overlap of the wave functions involved in the transition.

\section{Summary and conclusions}

We have presented an extended overview on the electronic and optical properties of different kinds of organic materials computed from first principles.
Adopting a solid-state physicists' perspective, we have described the systems in various states of matter, ranging from isolated molecules in gas-phase and in solution to extended crystals.
With the example of anthracene, we have shown that accounting for the native periodicity of the material is particularly relevant in order to capture the \textbf{k}-dispersion of the electronic bands and to provide a reliable starting point for the calculation of the optical properties, including excitons.
On the other hand, mimicking the crystalline environment by immersing a single molecule in a uniform electrostatic cavity, as enabled by the PCM, leads to very accurate values for the fundamental gap as well as for electron affinity and ionization potential.
This result suggests that when these properties are sought, an effective approach, considerably cheaper in terms of computational efforts than the full-fledged simulations of the whole crystal, can be successfully employed.
Similarly, adopting \textsc{LayerPCM}, the extension of PCM to deal with anisotropic layered substrates that we recently introduced~\cite{krum+21jcp}, grants access to the correct band-gap renormalization of conjugated molecules adsorbed on a TMDC monolayer.
The application of this method to other low-dimensional, polarizable substrates is straightforward.
Finally, as the last example, we considered a a charge-transfer complex formed by $p$-doped oligothiophene and inspected the mutual influence of short- and long-range interactions on its electronic and optical properties by simulating the system both as an isolated cluster and as a periodic co-crystal.
We found that only accounting for the lattice periodicity, one can correctly reproduce the intrinsic Bloch character of the wave function in the lattice, which crucially impact on the electronic and optical properties of the system.
These effects can be rationalized with the aid of a tight-binding model in the basis of the donor/acceptor unit.
This approach not only unfolds the physics of the problem in a transparent manner but, most importantly, draws a connection to consolidated theoretical methods for the description, for example, of the transport properties in this class of systems~\cite{groz+08irpc,troi11csr,ortm+11pssb}.

In our analysis, we have focused on the strengths of established solid-state physics approaches such as DFT, alone and coupled with effective models such as PCM and \textsc{LayerPCM}, and MBPT. 
Of course, there are many aspects related to the properties of organic materials that cannot be quantitatively captured with these methods.
For example, exciton dissociation and the subsequent dynamics of the photogenerated charge-carriers can be hardly described at the mean-field level of DFT and even MBPT cannot directly deliver more information than exciton distribution and binding energies.
Model Hamiltonians accounting for all relevant interactions are much for suitable to accomplish these tasks~\cite{altm+22jpcl,mikh+15ees,enge-enge17pccp}.
Likewise, intersections with machine learning have led to steps forward in crystal structure prediction of organic materials~\cite{hoer+19cpc} even with targeted properties, such as enhanced singlet fission~\cite{liu+22npjcm}.
These few examples shows the importance of establishing a common ground for truly interdisciplinary research in the field of organic materials. 
By offering our view on this topic, we hope to have provided our contribution in this direction.


\section*{Acknowledgement}
This work was funded by the German Federal Ministry of Education and Research (Professorinnenprogramm III) as well as from the Lower Saxony State (Professorinnen für Niedersachsen) and by the Deutsche Forschungsgemeinschaft (DFG) - Projektnummer 182087777 - SFB 951.
Computational resources were provided by the North-German Supercomputing Alliance (HLRN), project bep00076, and by the high-performance computing cluster CARL at the University of Oldenburg, funded by the German Research Foundation (Project No. INST 184/157-1 FUGG) and by the Ministry of Science and Culture of the Lower Saxony State.


\section*{Data Availability Statement}
The data that support the findings of this study are openly available on Zenodo at the following links: 10.5281/zenodo.7334160 (electronic and optical properties of anthracene molecules crystal); 10.5281/zenodo.7333310 (electronic and optical properties of donor/acceptor co-crystals); 10.5281/zenodo.7329775 (molecules on implicit substrates).

\section*{Appendix: Computational Details}\label{sec:comput}

\subsection*{Anthracene molecule and crystal}\label{ssec:ANT}
The results reported in Section~\ref{sec:ant} for anthracene molecules and crystals are obtained through a two-step procedure.
First, all structures are optimized by minimization of the interatomic forces until the threshold of 10$^{-3}$ eV/\AA{}. In the optimization of the molecular clusters, the positions of C atoms are clamped in order to preserve the herringbone angle between the two molecules assigned in the input structure to mimic their arrangement in the crystal.
Otherwise, the dimer relaxes to a configuration with the basal planes of the two molecules facing each other. These calculations are performed with the all-electron code FHI-aims~\cite{blum+09cpc} adopting tight integration grids, TIER2 basis sets~\cite{havu+09jcp}, and the generalized-gradient approximation in the Perdew-Burke-Ernzerhof (PBE) parameterization~\cite{perd+96prl} for the exchange-correlation potential.
Van der Waals interactions are included via the pairwise Tkatchenko-Scheffler scheme~\cite{tkat-sche09prl}. 
 
In the second step, electronic and optical properties are computed for the obtained geometries. 
For the finite systems (isolated molecules and clusters), the \textsc{MOLGW} code~\cite{brun+16cpc} is employed with Gaussian-type cc-pVTZ basis sets~\cite{brun12jcp} in conjunction with the frozen-core approximation and the resolution-of-identity approximation~\cite{weig+02}. 
The range-separated hybrid functional CAM-B3LYP~\cite{yana+04cpl} is adopted in the DFT step as a starting point for the subsequent $G_0W_0$ and BSE runs.
Calculations on the crystal are performed using the all-electron full-potential code \texttt{exciting}, implementing the family of linearized augmented plane-wave plus local orbitals (LAPW+LO) methods~\cite{gula+14jpcm}. 
Muffin-tin radii of 1.05~bohr and 0.7~bohr are chosen for carbon and hydrogen, respectively, along with a plane-wave cutoff G$_{max}$=4.7~bohr$^{-1}$.
The GGA in the PBE parameterization~\cite{perd+96prl} is used for the exchange-correlation potential.
In the DFT calculations, the BZ is sampled by a $8\times6\times5$ \textbf{k}-mesh, while in the BSE step~\cite{vorw+19es} a $\Gamma$-shifted $8\times6\times5$ \textbf{k}-point mesh is adopted.  
The QP correction is added to the KS gap through a scissors operator equal to $\delta =$2.27~eV, extimated by aligning the lowest-energy excitation computed for the ANT crystal with available experimental data~\cite{geac-pope69jcp}.
In the band-structure plot shown in Fig.~\ref{fig.ant.el}c), the binding energy of the first exciton, $E_b =$1.30~eV, output by our BSE calculation, has been added to $\delta$, in analogy with the procedure adopted in Ref.~\cite{cocc+18pccp}.
In the solution of the BSE, the screened Coulomb potential $W$ is computed using 200 empty bands and the energy threshold for the local field effects is set to 1.0~Ha.
In the construction and diagonalization of the BSE Hamiltonian, 10 occupied and 10 unoccupied bands are considered.

\subsection*{DFT+PCM and \textsc{LayerPCM}}

DFT calculations coupled with the PCM are performed with a locally modified copy of the \textsc{Octopus} code (v9.2)~\cite{tanc+2020jcp}. Nuclear potentials and core electrons are approximated by norm-conserving SG15 pseudopotentials~\cite{schl+gygi2015cpc}, the exchange-correlation potential by PBE~\cite{perd+96prl}. Neutral and singly-ionized molecules are treated in spin-restricted and unrestricted frameworks, respectively. KS wave functions are represented on a real-space grid generated by uniform sampling with spacing 0.21~\AA{} of a union of spheres of radius 4.5~\AA{} centered at each atomic position. The PCM cavity is generated in a similar, yet not identical fashion~\cite{pasc+sill1990jcp}, using smaller spheres with atom-dependent radii. Here, we take the van der Waals radius of the respective element, scaled by a factor of 1.1. 

For the atomistic calculation of the isolated TMDC monolayer and the ANT@MoS$_2$ interface as well as for the determination of the dielectric constants fed into the PCM, we exploit the interface between the \textsc{Quantum Espresso} suite (v6.8)~\cite{gian+09jpcm} and the \textsc{Yambo} code (v5.0.2)~\cite{marini2009,Sangalli_jpcm}. With the former, we obtain the DFT starting point for many-body calculations performed with the latter. We again employ SG15 pseudopotentials~\cite{schl+gygi2015cpc} for the cores and PBE for the exchange-correlation functional~\cite{perd+96prl}, neglecting spin-orbit coupling. Structural relaxations are done with a wave-function cutoff of 60~Ry and including the pairwise Tkatschenko-Scheffler dispersion correction~\cite{tkat-sche09prl}, reducing the components of all forces to below 10$^{-3}$ Ha/bohr. At this step, we use \textbf{k}-point samplings of $8\times8\times1$, $2\times2\times1$, and $4\times6\times4$ for MoS$_2$, ANT@MoS$_2$, and the acene crystals, respectively.
To simulate the ANT@MoS$_2$ interface, we choose a lattice constant of 20~\AA{} in the non-periodic direction including a sufficiently large vacuum region to decouple the replicas, and we include a dipole correction. We use Wannier interpolation \cite{marz+12rmr} as implemented in the \textsc{Wannier90} code (v3.1.0) \cite{pizz+2020jpcm} to determine the density of states of ANT@MoS$_2$ on a $100\times100\times1$ \textbf{k}-grid, allowing for the application of only minimal artificial broadening in the electronic structure, which endows the obtained linewidths with physical meaningfulness. The corresponding band structure is unfolded to the BZ of the MoS$_2$ unit cell~\cite{krum-cocc21es,popescuZunger2012prb}. The DFT calculations for the generation of the required large number of unoccupied orbitals in the many-body simulations are done with a wavefunction cutoff of 40~Ry, sticking with the same $4\times6\times4$ \textbf{k}-mesh for the acenes, but increasing it to $24\times24\times1$ and $6\times6\times1$ for MoS$_2$ and ANT@MoS$_2$, respectively, which is necessary to obtain converged $GW$ results. We calculate the dielectric functions of the acenes by means of linear-response calculation on the level of the RPA, featuring 300 conduction bands and a $\textbf{G},\textbf{G}'$ sum kinetic cutoff of 3~Ry. The values used in the PCM correspond to the trace of the $\omega=0$ dielectric tensor, averaging $xx$, $yy$, and $zz$ components. For MoS$_2$, we take 100 conduction bands and a cutoff of 5~Ry to calculate the (effective) anisotropic dielectric functions within the RPA.  

The $GW$ calculations on MoS$_2$ and ANT@MoS$_2$ are performed with the \textsc{Yambo} code~\cite{marini2009,Sangalli_jpcm} adopting the plasmon-pole approximation for the dynamically screened interaction $W$. 300 (MoS$_2$) and 500 (ANT@MoS$_2$) conduction bands are included for both $W$ and the correlation part of the self-energy, $\Sigma_c$ and employ the sum-over-states terminator~\cite{brun+gonz08prb}. 
The $\textbf{G},\textbf{G}'$ sum is cut off at kinetic energies of 5~Ry. A Coulomb cutoff~\cite{rozz+07prb} along the out-of-plane direction is applied, truncating the interaction at 19.57\,\AA\,$\lesssim$\,($c=20$\,\AA). Coulomb integrals over the BZ for the smallest $\sim$300 \textbf{G} vectors are calculated with the random integration method, replacing the sum over a uniformly sampled Brillouin zone by a 10$^6$ \textbf{q}-point Monte-Carlo integration. Here, we neglect the anisotropy of $W$ for small $\textbf{q}$, choosing the corresponding polarization vector as $\frac{(1,1,1)}{\sqrt{3}}$, representing an orientational average.


\subsection*{Donor/acceptor complex and co-crystal}
The results for the 4T:TCNQ co-crystal presented in Section~\ref{sec:co-cry} have been obtained from DFT using the code \textsc{Quantum Espresso}~\cite{gian+09jpcm} and from MBPT using \textsc{Yambo}~\cite{marini2009,Sangalli_jpcm}. 
In the former step, a plane-wave basis set cutoff of 50~Ry (200~Ry) for the wave-functions (electron density) and norm conserving pseudopotentials~\cite{Hamann_prb} are employed. 
The PBE functional~\cite{perd+96prl} is adopted in conjunction with the Tkatchenko-Scheffler pairwise scheme~\cite{tkat-sche09prl} to account for van der Waals interactions. 
A 4$\times$4$\times$4 \textbf{k}-grid is used to sample the BZ.
Self-consistent calculations are carried out on optimized geometries obtained by minimizing interatomic forces with a threshold of 10$^{-4}$~Ry/\AA{}.
BSE calculations are performed on top of the DFT electronic structure with the quasi-particle correction added in the form of a scissors operator of 1.41~eV to the conduction bands based on available experimental spectra~\cite{mend+15ncom}. 
The BSE Hamiltonian is constructed including 20 occupied and 40 unoccupied states, and by 4$\times$4$\times$4 \textbf{k}-point grid; it is diagonalized using the Haydock-Lanczos algorithm~\cite{Lanczos1950}.

Calculations on the 4T:TCNQ cluster are performed with \textsc{MOLGW}~\cite{brun+16cpc} adopting the same computational parameters employed for ANT, except for the Gaussian basis set, which is here reduced to aug-cc-pVDZ.

\section*{References}

\begin{thebibliography}{100}
\expandafter\ifx\csname url\endcsname\relax
  \def\url#1{{\tt #1}}\fi
\expandafter\ifx\csname urlprefix\endcsname\relax\def\urlprefix{URL }\fi
\providecommand{\eprint}[2][]{\url{#2}}

\bibitem{brue05book}
Br\"utting W 2005 {\em Introduction to the Physics of Organic Semiconductors\/}
  (John Wiley \& Sons, Ltd) pp 1--14 ISBN 9783527606634

\bibitem{myer-xue12}
Myers J~D and Xue J 2012 {\em Polym.~Rev.\/} {\bf 52} 1--37

\bibitem{horo98am}
Horowitz G 1998 {\em Adv.~Mater.\/} {\bf 10} 365--377

\bibitem{pron-rann02pps}
Pron A and Rannou P 2002 {\em Prog.~Poli.~Sci.\/} {\bf 27} 135--190

\bibitem{sing-sari06armr}
Singh T~B and Sariciftci N~S 2006 {\em Ann.~Rev.~Mat.~Res.\/} {\bf 36} 199--230

\bibitem{shir-loui93prl}
Shirley E~L and Louie S~G 1993 {\em Phys.~Rev.~Lett.\/} {\bf 71}(1) 133--136
  \urlprefix\url{https://link.aps.org/doi/10.1103/PhysRevLett.71.133}

\bibitem{ruin+02prl}
Ruini A, Caldas M~J, Bussi G and Molinari E 2002 {\em Phys.~Rev.~Lett.\/} {\bf
  88} 206403

\bibitem{buss+02apl}
Bussi G, Ruini A, Molinari E, Caldas M~J, Puschnig P and Ambrosch-Draxl C 2002
  {\em Appl.~Phys.~Lett.\/} {\bf 80} 4118--4120

\bibitem{pusc+02prl}
Puschnig P and Ambrosch-Draxl C 2002 {\em Phys.~Rev.~Lett.\/} {\bf 89} 056405

\bibitem{tiag+03prb}
Tiago M~L, Northrup J~E and Louie S~G 2003 {\em Phys.~Rev.~B\/} {\bf 67} 115212

\bibitem{humm+04prl}
Hummer K, Puschnig P and Ambrosch-Draxl C 2004 {\em Phys.~Rev.~Lett.\/} {\bf
  92}(14) 147402
  \urlprefix\url{https://link.aps.org/doi/10.1103/PhysRevLett.92.147402}

\bibitem{humm-ambr05prb}
Hummer K and Ambrosch-Draxl C 2005 {\em Phys.~Rev.~B\/} {\bf 71} 081202

\bibitem{humm-drax05prb2}
Hummer K and Ambrosch-Draxl C 2005 {\em Phys.~Rev.~B\/} {\bf 72} 205205

\bibitem{hahn+05prb}
Hahn P~H, Schmidt W~G and Bechstedt F 2005 {\em Phys.~Rev.~B\/} {\bf 72}(24)
  245425 \urlprefix\url{https://link.aps.org/doi/10.1103/PhysRevB.72.245425}

\bibitem{henn+99cp}
Hennessy M, Soos Z, Pascal~Jr R and Girlando A 1999 {\em Chem.~Phys.\/} {\bf
  245} 199--212

\bibitem{casi+02prb}
Casian A, Dusciac V and Coropceanu I 2002 {\em Phys.~Rev.~B\/} {\bf 66} 165404

\bibitem{hoff-soos02prb}
Hoffmann M and Soos Z 2002 {\em Phys.~Rev.~B\/} {\bf 66} 024305

\bibitem{hann-bobb04prb}
Hannewald K and Bobbert P 2004 {\em Phys.~Rev.~B\/} {\bf 69} 075212

\bibitem{hann+04prb}
Hannewald K, Stojanovi\ifmmode~\acute{c}\else \'{c}\fi{} V~M, Schellekens
  J~M~T, Bobbert P~A, Kresse G and Hafner J 2004 {\em Phys.~Rev.~B\/} {\bf
  69}(7) 075211
  \urlprefix\url{https://link.aps.org/doi/10.1103/PhysRevB.69.075211}

\bibitem{kara-bitt04jpcb}
Karabunarliev S and Bittner E~R 2004 {\em J.~Phys.~Chem.~B\/} {\bf 108}
  10219--10225

\bibitem{bitt+05jcp}
Bittner E~R, Ramon J~G~S and Karabunarliev S 2005 {\em J.~Chem.~Phys.\/} {\bf
  122} 214719

\bibitem{troi+06prl}
Troisi A and Orlandi G 2006 {\em Phys.~Rev.~Lett.\/} {\bf 96} 086601

\bibitem{cram-thom01jpca}
Cramer C~J and Thompson J 2001 {\em J.~Phys.~Chem.~A\/} {\bf 105} 2091--2098

\bibitem{tsuz+02jacs}
Tsuzuki S, Honda K and Azumi R 2002 {\em J.~Am.~Chem.~Soc.\/} {\bf 124}
  12200--12209

\bibitem{naka+03jpca}
Nakano M, Yamada S, Takahata M and Yamaguchi K 2003 {\em J.~Phys.~Chem.~A\/}
  {\bf 107} 4157--4164

\bibitem{you+04cpl}
You Z~Q, Shao Y and Hsu C~P 2004 {\em Chem.~Phys.~Lett.\/} {\bf 390} 116--123

\bibitem{hutc+02jpca}
Hutchison G~R, Ratner M~A and Marks T~J 2002 {\em J.~Phys.~Chem.~A\/} {\bf 106}
  10596--10605

\bibitem{weib-yaro02jcp}
Weibel J~D and Yaron D 2002 {\em J.~Chem.~Phys.\/} {\bf 116} 6846--6856

\bibitem{risk+04jcp}
Risko C, Kushto G, Kafati Z and Br{\'e}das J~L 2004 {\em J.~Chem.~Phys.\/} {\bf
  121} 9031--9038

\bibitem{chho+16nrm}
Chhowalla M, Jena D and Zhang H 2016 {\em Nat.~Rev.~Mater.\/} {\bf 1} 1--15

\bibitem{bren+16nrm}
Brenner T~M, Egger D~A, Kronik L, Hodes G and Cahen D 2016 {\em
  Nat.~Rev.~Mater.\/} {\bf 1} 1--16

\bibitem{umar+14sr}
Umari P, Mosconi E and De~Angelis F 2014  {\bf 4} 1--7

\bibitem{fili-gius14prb}
Filip M~R and Giustino F 2014 {\em Phys.~Rev.~B\/} {\bf 90} 245145

\bibitem{briv+14prb}
Brivio F, Butler K~T, Walsh A and Van~Schilfgaarde M 2014 {\em Phys.~Rev.~B\/}
  {\bf 89} 155204

\bibitem{bokd+16sr}
Bokdam M, Sander T, Stroppa A, Picozzi S, Sarma D, Franchini C and Kresse G
  2016  {\bf 6} 1--8

\bibitem{mott+15natcom}
Motta C, El-Mellouhi F, Kais S, Tabet N, Alharbi F and Sanvito S 2015 {\em
  Nature Commun.\/} {\bf 6} 1--7

\bibitem{he+18acsel}
He J, Fang W~H, Long R and Prezhdo O~V 2018 {\em ACS~Energy~Lett.\/} {\bf 3}
  2070--2076

\bibitem{ghos+20jpcl}
Ghosh D, Welch E, Neukirch A~J, Zakhidov A and Tretiak S 2020 {\em
  J.~Phys.~Chem.~Lett.\/} {\bf 11} 3271--3286

\bibitem{li+22jpcc}
Li S, Zhao S, Chu H, Gao Y, Lv P, Wang V, Tang G and Hong J 2022 {\em
  J.~Phys.~Chem.~C\/} {\bf 126} 4715--4725

\bibitem{fili+21prl}
Filip M~R, Haber J~B and Neaton J~B 2021 {\em Phys.~Rev.~Lett.\/} {\bf 127}
  067401

\bibitem{zhu+11cm}
Zhu L, Kim E~G, Yi Y and Br\'{e}das J~L 2011 {\em Chem.~Mater.\/} {\bf 23}
  5149--5159

\bibitem{salz+13prl}
Salzmann I, Heimel G, Duhm S, Oehzelt M, Pingel P, George B~M, Schnegg A, Lips
  K, Blum R~P, Vollmer A and Koch N 2012 {\em Phys.~Rev.~Lett.\/} {\bf 108}(3)
  035502
  \urlprefix\url{https://link.aps.org/doi/10.1103/PhysRevLett.108.035502}

\bibitem{zhu+14jpcc}
Zhu L, Yi Y, Fonari A, Corbin N~S, Coropceanu V and Bredas J~L 2014 {\em
  J.~Phys.~Chem.~C\/} {\bf 118} 14150--14156

\bibitem{mend+15ncom}
M\'{e}ndez H, Heimel G, Winkler S, Frisch J, Opitz A, Sauer K, Wegner B,
  Oehzelt M, R\"{o}thel C, Duhm S, T\"{o}bbens D, Koch N and Salzmann I 2015
  {\em Nature Commun.\/} {\bf 6} 8560

\bibitem{salz+16acr}
Salzmann I, Heimel G, Oehzelt M, Winkler S and Koch N 2016 {\em
  Acc.~Chem.~Res.\/} {\bf 49} 370--378

\bibitem{li+17prm}
Li J, D'Avino G, Pershin A, Jacquemin D, Duchemin I, Beljonne D and Blase X
  2017  {\bf 1}(2) 025602
  \urlprefix\url{https://link.aps.org/doi/10.1103/PhysRevMaterials.1.025602}

\bibitem{beye+19cm}
Beyer P, Pham D, Peter C, Koch N, Meister E, Brutting W, Gr\"{u}bert L, Hecht
  S, Nabok D, Cocchi C, Draxl C and Opitz A 2019 {\em Chem.~Mater.\/} {\bf 31}
  1237--1249 (\textit{Preprint}
  \eprint{https://doi.org/10.1021/acs.chemmater.8b01447})
  \urlprefix\url{https://doi.org/10.1021/acs.chemmater.8b01447}

\bibitem{ambr+09njp}
Ambrosch-Draxl C, Nabok D, Puschnig P and Meisenbichler C 2009 {\em
  New.~J.~Phys.\/} {\bf 11} 125010

\bibitem{pith+15cgd}
Pithan L, Cocchi C, Zschiesche H, Weber C, Zykov A, Bommel S, Leake S~J,
  Schäfer P, Draxl C and Kowarik S 2015 {\em Cryst.~Growth~Des.\/} {\bf 15}
  1319--1324

\bibitem{klet+16pccp}
Klett B, Cocchi C, Pithan L, Kowarik S and Draxl C 2016 {\em
  Phys.~Chem.~Chem.~Phys.\/} {\bf 18}(21) 14603--14609

\bibitem{jaco+18mh}
Jacobs I~E, Cendra C, Harrelson T~F, Bedolla~Valdez Z~I, Faller R, Salleo A and
  Moul\'{e} A~J 2018 {\em Mater.~Horiz.\/} {\bf 5}(4) 655--660
  \urlprefix\url{http://dx.doi.org/10.1039/C8MH00223A}

\bibitem{cocc+18pccp}
Cocchi C, Breuer T, Witte G and Draxl C 2018 {\em Phys.~Chem.~Chem.~Phys.\/}
  {\bf 20} 29724--29736

\bibitem{zimm+11jacs}
Zimmerman P~M, Bell F, Casanova D and Head-Gordon M 2011 {\em
  J.~Am.~Chem.~Soc.\/} {\bf 133} 19944--19952

\bibitem{shar+13jpcl}
Sharifzadeh S, Darancet P, Kronik L and Neaton J~B 2013 {\em
  J.~Phys.~Chem.~Lett.\/} {\bf 4} 2197--2201

\bibitem{berk+14jcp}
Berkelbach T~C, Hybertsen M~S and Reichman D~R 2014 {\em J.~Chem.~Phys.\/} {\bf
  141} 074705

\bibitem{refa+17prl}
Refaely-Abramson S, da~Jornada F~H, Louie S~G and Neaton J~B 2017 {\em
  Phys.~Rev.~Lett.\/} {\bf 119}(26) 267401
  \urlprefix\url{https://link.aps.org/doi/10.1103/PhysRevLett.119.267401}

\bibitem{wang+18jcp}
Wang X, Liu X, Cook C, Schatschneider B and Marom N 2018 {\em J.~Chem.~Phys.\/}
  {\bf 148} 184101

\bibitem{abra-mayh21jpcl}
Abraham V and Mayhall N~J 2021 {\em J.~Phys.~Chem.~Lett.\/} {\bf 12}
  10505--10514

\bibitem{altm+22jpcl}
Altman A~R, Refaely-Abramson S and da~Jornada F~H 2022 {\em
  J.~Phys.~Chem.~Lett.\/} {\bf 13} 747--753

\bibitem{cuda+12prb}
Cudazzo P, Gatti M and Rubio A 2012 {\em Phys.~Rev.~B\/} {\bf 86} 195307

\bibitem{rang+16prb}
Rangel T, Berland K, Sharifzadeh S, Brown-Altvater F, Lee K, Hyldgaard P,
  Kronik L and Neaton J~B 2016 {\em Phys.~Rev.~B\/} {\bf 93} 115206

\bibitem{zhu+19jpcl}
Zhu L, Tu Z, Yi Y and Wei Z 2019 {\em J.~Phys.~Chem.~Lett.\/} {\bf 10}
  4888--4894

\bibitem{sun+16jctc}
Sun H, Ryno S, Zhong C, Ravva M~K, Sun Z, K\"orzd\"orfer T and Bredas J~L 2016
  {\em J.~Chem.~Theory.~Comput.\/} {\bf 12} 2906--2916

\bibitem{cast+17cpl}
Castro A~N, Osorio F~A, Ternavisk R~R, Napolitano H~B, Valverde C and Baseia B
  2017 {\em Chem.~Phys.~Lett.\/} {\bf 681} 110--123

\bibitem{duch+18cs}
Duchemin I, Guido C~A, Jacquemin D and Blase X 2018 {\em Chem.~Sci.\/} {\bf 9}
  4430--4443

\bibitem{ross-sohl09jpcc}
Rossi M and Sohlberg K 2009 {\em J.~Phys.~Chem.~C\/} {\bf 113} 6821--6831

\bibitem{brow+20prb}
Brown-Altvater F, Antonius G, Rangel T, Giantomassi M, Draxl C, Gonze X, Louie
  S~G and Neaton J~B 2020 {\em Phys.~Rev.~B\/} {\bf 101}(16) 165102
  \urlprefix\url{https://link.aps.org/doi/10.1103/PhysRevB.101.165102}

\bibitem{cook-bera20jcp}
Cook C and Beran G~J~O 2020 {\em J.~Chem.~Phys.\/} {\bf 153} 224105

\bibitem{jacq+09jctc}
Jacquemin D, Wathelet V, Perpete E~A and Adamo C 2009 {\em
  J.~Chem.~Theory.~Comput.\/} {\bf 5} 2420--2435

\bibitem{shar+12prb}
Sharifzadeh S, Biller A, Kronik L and Neaton J~B 2012 {\em Phys.~Rev.~B\/} {\bf
  85}(12) 125307
  \urlprefix\url{https://link.aps.org/doi/10.1103/PhysRevB.85.125307}

\bibitem{hohe-kohn64pr}
Hohenberg P and Kohn W 1964 {\em Phys. Rev.\/} {\bf 136}(3B) B864--B871
  \urlprefix\url{https://link.aps.org/doi/10.1103/PhysRev.136.B864}

\bibitem{kohn-sham65pr}
Kohn W and Sham L~J 1965 {\em Phys.~Rev.\/} {\bf 140}(4A) A1133--A1138
  \urlprefix\url{https://link.aps.org/doi/10.1103/PhysRev.140.A1133}

\bibitem{perd-schm01aipcp}
Perdew J~P and Schmidt K 2001 {\em AIP~Conf.~Proc.\/} {\bf 577} 1--20

\bibitem{perd86prb}
Perdew J~P 1986 {\em Phys.~Rev.~B\/} {\bf 33}(12) 8822--8824
  \urlprefix\url{https://link.aps.org/doi/10.1103/PhysRevB.33.8822}

\bibitem{tran-blah09prl}
Tran F and Blaha P 2009 {\em Phys.~Rev.~Lett.\/} {\bf 102}(22) 226401

\bibitem{sun+15prl}
Sun J, Ruzsinszky A and Perdew J~P 2015 {\em Phys. Rev. Lett.\/} {\bf 115}(3)
  036402
  \urlprefix\url{https://link.aps.org/doi/10.1103/PhysRevLett.115.036402}

\bibitem{beck93jcp}
Becke A~D 1993 {\em J.~Chem.~Phys.\/} {\bf 98} 1372--1377

\bibitem{iiku+01jcp}
Iikura H, Tsuneda T, Yanai T and Hirao K 2001 {\em J.~Chem.~Phys.\/} {\bf 115}
  3540--3544

\bibitem{toul+04pra}
Toulouse J, Colonna F~m~c and Savin A 2004 {\em Phys.~Rev.~A\/} {\bf 70}(6)
  062505 \urlprefix\url{https://link.aps.org/doi/10.1103/PhysRevA.70.062505}

\bibitem{hse03}
Heyd J, Scuseria G~E and Ernzerhof M 2003 {\em J.~Chem.~Phys.\/} {\bf 118}
  8207--8215

\bibitem{gula+14jpcm}
Gulans A, Kontur S, Meisenbichler C, Nabok D, Pavone P, Rigamonti S, Sagmeister
  S, Werner U and Draxl C 2014 {\em J.~Phys.:~Condens.~Matter.\/} {\bf 26}
  363202 \urlprefix\url{https://doi.org/10.1088/0953-8984/26/36/363202}

\bibitem{gian+17jpcm}
Giannozzi P, Andreussi O, Brumme T, Bunau O, Nardelli M~B, Calandra M, Car R,
  Cavazzoni C, Ceresoli D, Cococcioni M, Colonna N, Carnimeo I, Corso A~D,
  de~Gironcoli S, Delugas P, DiStasio R~A, Ferretti A, Floris A, Fratesi G,
  Fugallo G, Gebauer R, Gerstmann U, Giustino F, Gorni T, Jia J, Kawamura M, Ko
  H~Y, Kokalj A, K\"{u}{\c{c}}\"{u}kbenli E, Lazzeri M, Marsili M, Marzari N,
  Mauri F, Nguyen N~L, Nguyen H~V, de-la Roza A~O, Paulatto L, Ponc{\'{e}} S,
  Rocca D, Sabatini R, Santra B, Schlipf M, Seitsonen A~P, Smogunov A, Timrov
  I, Thonhauser T, Umari P, Vast N, Wu X and Baroni S 2017 {\em
  J.~Phys.:~Condens.~Matter.\/} {\bf 29} 465901

\bibitem{g16}
Frisch M~J, Trucks G~W, Schlegel H~B, Scuseria G~E, Robb M~A, Cheeseman J~R,
  Scalmani G, Barone V, Petersson G~A, Nakatsuji H, Li X, Caricato M, Marenich
  A~V, Bloino J, Janesko B~G, Gomperts R, Mennucci B, Hratchian H~P, Ortiz J~V,
  Izmaylov A~F, Sonnenberg J~L, Williams-Young D, Ding F, Lipparini F, Egidi F,
  Goings J, Peng B, Petrone A, Henderson T, Ranasinghe D, Zakrzewski V~G, Gao
  J, Rega N, Zheng G, Liang W, Hada M, Ehara M, Toyota K, Fukuda R, Hasegawa J,
  Ishida M, Nakajima T, Honda Y, Kitao O, Nakai H, Vreven T, Throssell K,
  Montgomery {Jr} J~A, Peralta J~E, Ogliaro F, Bearpark M~J, Heyd J~J, Brothers
  E~N, Kudin K~N, Staroverov V~N, Keith T~A, Kobayashi R, Normand J,
  Raghavachari K, Rendell A~P, Burant J~C, Iyengar S~S, Tomasi J, Cossi M,
  Millam J~M, Klene M, Adamo C, Cammi R, Ochterski J~W, Martin R~L, Morokuma K,
  Farkas O, Foresman J~B and Fox D~J 2016 Gaussian 16 revision a.03 gaussian
  Inc. Wallingford CT

\bibitem{blum+09cpc}
Blum V, Gehrke R, Hanke F, Havu P, Havu V, Ren X, Reuter K and Scheffler M 2009
  {\em Comput.~Phys.~Commun.\/} {\bf 180} 2175 -- 2196

\bibitem{andr+15pccp}
Andrade X, Strubbe D, De~Giovannini U, Larsen A~H, Oliveira M~J~T,
  Alberdi-Rodriguez J, Varas A, Theophilou I, Helbig N, Verstraete M~J, Stella
  L, Nogueira F, Aspuru-Guzik A, Castro A, Marques M~A~L and Rubio A 2015 {\em
  Phys.~Chem.~Chem.~Phys.\/} {\bf 17}(47) 31371--31396
  \urlprefix\url{http://dx.doi.org/10.1039/C5CP00351B}

\bibitem{rung-gros84prl}
Runge E and Gross E~K~U 1984 {\em Phys.~Rev.~Lett.\/} {\bf 52}(12) 997--1000

\bibitem{casi+98jcp}
Casida M, Jamorski C, Casida K and Salahub D 1998 {\em J.~Chem.~Phys.\/} {\bf
  108} 4439--4449

\bibitem{marq-gros04arpc}
Marques M~A and Gross E~K 2004 {\em Annu.~Rev.~Phys.~Chem.\/} {\bf 55} 427--455

\bibitem{cocc+14prl}
Cocchi C, Prezzi D, Ruini A, Molinari E and Rozzi C~A 2014 {\em
  Phys.~Rev.~Lett.\/} {\bf 112}(19) 198303

\bibitem{guan+21pccp}
Guandalini A, Cocchi C, Pittalis S, Ruini A and Rozzi C~A 2021 {\em
  Phys.~Chem.~Chem.~Phys.\/} {\bf 23} 10059--10069

\bibitem{degi+13chpch}
De~Giovannini U, Brunetto G, Castro A, Walkenhorst J and Rubio A 2013 {\em
  Chem.~Phys.~Chem.\/} {\bf 14} 1363--1376

\bibitem{krum+20jcp}
Krumland J, Valencia A~M, Pittalis S, Rozzi C~A and Cocchi C 2020 {\em
  J.~Chem.~Phys.\/} {\bf 153} 054106

\bibitem{neug+05jcp}
Neugebauer J, Louwerse M~J, Baerends E~J and Wesolowski T~A 2005 {\em
  J.~Chem.~Phys.\/} {\bf 122} 094115

\bibitem{bott+07rpp}
Botti S, Schindlmayr A, Del~Sole R and Reining L 2007 {\em Rep.~Prog.~Phys.\/}
  {\bf 70} 357

\bibitem{sott+05ijqc}
Sottile F, Bruneval F, Marinopoulos A, Dash L, Botti S, Olevano V, Vast N,
  Rubio A and Reining L 2005 {\em Int.~J.~Quantum~Chem.\/} {\bf 102} 684--701

\bibitem{cocc-drax15prb}
Cocchi C and Draxl C 2015 {\em Phys.~Rev.~B\/} {\bf 92}(20) 205126
  \urlprefix\url{https://link.aps.org/doi/10.1103/PhysRevB.92.205126}

\bibitem{onid+02rmp}
Onida G, Reining L and Rubio A 2002 {\em Rev.~Mod.~Phys.\/} {\bf 74}(2)
  601--659 \urlprefix\url{https://link.aps.org/doi/10.1103/RevModPhys.74.601}

\bibitem{hedi65pr}
Hedin L 1965 {\em Phys.~Rev.\/} {\bf 139}(3A) A796--A823
  \urlprefix\url{https://link.aps.org/doi/10.1103/PhysRev.139.A796}

\bibitem{salp-beth51pr}
Salpeter E~E and Bethe H~A 1951 {\em Phys.~Rev.\/} {\bf 84} 1232

\bibitem{hank-sham80prb}
Hanke W and Sham L~J 1980 {\em Phys.~Rev.~B\/} {\bf 21}(10) 4656--4673
  \urlprefix\url{https://link.aps.org/doi/10.1103/PhysRevB.21.4656}

\bibitem{hybe-loui85prl}
Hybertsen M~S and Louie S~G 1985 {\em Phys.~Rev.~Lett.\/} {\bf 55} 1418

\bibitem{marini2009}
Marini A, Hogan C, Gr\"{u}ning M and Varsano D 2009 {\em
  Comput.~Phys.~Commun.\/} {\bf 180} 1392 -- 1403 ISSN 0010-4655

\bibitem{vorw+19es}
Vorwerk C, Aurich B, Cocchi C and Draxl C 2019 {\em Electron.~Struct.\/} {\bf
  1} 037001

\bibitem{blas+11prb}
Blase X, Attaccalite C and Olevano V 2011 {\em Phys.~Rev.~B\/} {\bf 83}(11)
  115103 \urlprefix\url{https://link.aps.org/doi/10.1103/PhysRevB.83.115103}

\bibitem{hiro+15prb}
Hirose D, Noguchi Y and Sugino O 2015 {\em Phys.~Rev.~B\/} {\bf 91} 205111

\bibitem{brun+16cpc}
Bruneval F, Rangel T, Hamed S~M, Shao M, Yang C and Neaton J~B 2016 {\em
  Comput.~Phys.~Commun.\/} {\bf 208} 149 -- 161 ISSN 0010-4655
  \urlprefix\url{http://www.sciencedirect.com/science/article/pii/S0010465516301990}

\bibitem{rein18wircms}
Reining L 2018 {\em Wiley~Interdiscip.~Rev.~Comput.~Mol.~Sci.\/} {\bf 8} e1344

\bibitem{arya-gurn98rpp}
Aryasetiawan F and Gunnarsson O 1998 {\em Rep.~Prog.~Phys.\/} {\bf 61} 237

\bibitem{stri88rnc}
Strinati G 1988 {\em Riv.~Nuovo~Cimento\/} {\bf 11} 1--86 ISSN 1826-9850
  \urlprefix\url{https://doi.org/10.1007/BF02725962}

\bibitem{cocc+16jcp}
Cocchi C, Moldt T, Gahl C, Weinelt M and Draxl C 2016 {\em J.~Chem.~Phys.\/}
  {\bf 145} 234701

\bibitem{cocc-drax17jpcm}
Cocchi C and Draxl C 2017 {\em J.~Phys.:~Condens.~Matter.\/} {\bf 29} 394005

\bibitem{rohl-loui00prb}
Rohlfing M and Louie S~G 2000 {\em Phys.~Rev.~B\/} {\bf 62}(8) 4927--4944
  \urlprefix\url{https://link.aps.org/doi/10.1103/PhysRevB.62.4927}

\bibitem{fox02book}
Fox M 2002 Optical properties of solids

\bibitem{cocc-drax17jcpm}
Cocchi C and Draxl C 2017 {\em J.~Phys.:~Condens.~Matter.\/} {\bf 29} 394005

\bibitem{guer+21jpcc}
Guerrini M, Valencia A~M and Cocchi C 2021 {\em J.~Phys.~Chem.~C\/} {\bf 125}
  20821--20830

\bibitem{cocc+11jpcl}
Cocchi C, Prezzi D, Ruini A, Caldas M~J and Molinari E 2011 {\em
  J.~Phys.~Chem.~Lett.\/} {\bf 2} 1315--1319

\bibitem{deco+14jpcc}
De~Corato M, Cocchi C, Prezzi D, Caldas M~J, Molinari E and Ruini A 2014 {\em
  J.~Phys.~Chem.~C\/} {\bf 118} 23219--23225

\bibitem{toma+05cr}
Tomasi J, Mennucci B and Cammi R 2005 {\em Chem.~Rev.\/} {\bf 105} 2999--3094

\bibitem{cances1997jcp}
Cances E, Mennucci B and Tomasi J 1997 {\em J.~Chem.~Phys.\/} {\bf 107}
  3032--3041 (\textit{Preprint} \eprint{https://doi.org/10.1063/1.474659})
  \urlprefix\url{https://doi.org/10.1063/1.474659}

\bibitem{cances1998jmc}
Cancès E and Mennucci B 1998  {\bf 23} 309--326

\bibitem{corni2015jpca}
Corni S, Pipolo S and Cammi R 2015 {\em J.~Phys.~Chem.~A\/} {\bf 119}
  5405--5416 pMID: 25485456 (\textit{Preprint}
  \eprint{https://doi.org/10.1021/jp5106828})
  \urlprefix\url{https://doi.org/10.1021/jp5106828}

\bibitem{krum+21jcp}
Krumland J, Gil G, Corni S and Cocchi C 2021 {\em J.~Chem.~Phys.\/} {\bf 154}
  224114

\bibitem{prob-karl75pssa}
Probst K~H and Karl N 1975 {\em physica status solidi (a)\/} {\bf 27} 499--508
  (\textit{Preprint}
  \eprint{https://onlinelibrary.wiley.com/doi/pdf/10.1002/pssa.2210270219})
  \urlprefix\url{https://onlinelibrary.wiley.com/doi/abs/10.1002/pssa.2210270219}

\bibitem{math+50ac}
McL~Mathieson A, Robertson J~M and Sinclair V 1950 {\em
  Angew.~Chem.~Int.~Ed.\/} {\bf 3} 245--250

\bibitem{cuda+15jpcm}
Cudazzo P, Sottile F, Rubio A and Gatti M 2015 {\em
  J.~Phys.:~Condens.~Matter.\/} {\bf 27} 113204

\bibitem{Davydov_1964}
Davydov A~S 1964 {\em Soviet Physics Uspekhi\/} {\bf 7} 145--178
  \urlprefix\url{https://doi.org/10.1070/pu1964v007n02abeh003659}

\bibitem{refa+13prb}
Refaely-Abramson S, Sharifzadeh S, Jain M, Baer R, Neaton J~B and Kronik L 2013
  {\em Phys.~Rev.~B\/} {\bf 88} 081204

\bibitem{baes+kill1973mclc}
Baessler H and Killesreiter H 1973 {\em Mol.~Cryst.~Liq.~Cryst.\/} {\bf 24}
  21--31 (\textit{Preprint} \eprint{https://doi.org/10.1080/15421407308083385})
  \urlprefix\url{https://doi.org/10.1080/15421407308083385}

\bibitem{belk+grec1974pssa}
Belkind A~I and Grechov V~V 1974 {\em Phys.~Status~Solidi~A\/} {\bf 26} 377

\bibitem{riga+1977phsc}
Riga J, Pireaux J~J, Caudano R and Verbist J~J 1977 {\em Phys.~Scr.\/} {\bf 16}
  346--350 \urlprefix\url{https://doi.org/10.1088/0031-8949/16/5-6/026}

\bibitem{guen+21jpcl}
G\"under D, Valencia A~M, Guerrini M, Breuer T, Cocchi C and Witte G 2021 {\em
  J.~Phys.~Chem.~Lett.\/} {\bf 12} 9899--9905

\bibitem{zeis+21jpcl}
Zeiser C, Moretti L, Geiger T, Kalix L, Valencia A~M, Maiuri M, Cocchi C,
  Bettinger H~F, Cerullo G and Broch K 2021 {\em J.~Phys.~Chem.~Lett.\/} {\bf
  12} 7453--7458

\bibitem{stai+04jms}
Staicu A, Krasnokutski S and Rouill{\'e} G 2004 {\em J.~Mol.~Struct.\/} {\bf
  786} 105

\bibitem{vorw+16cpc}
Vorwerk C, Cocchi C and Draxl C 2016 {\em Comput.~Phys.~Commun.\/} {\bf 201}
  119--125

\bibitem{schmidt1977jcp}
Schmidt W 1977 {\em The Journal of Chemical Physics\/} {\bf 66} 828--845
  (\textit{Preprint} \eprint{https://doi.org/10.1063/1.433961})
  \urlprefix\url{https://doi.org/10.1063/1.433961}

\bibitem{boschi1974jcp}
Boschi R, Clar E and Schmidt W 1974 {\em J.~Chem.~Phys.\/} {\bf 60} 4406--4418
  (\textit{Preprint} \eprint{https://doi.org/10.1063/1.1680919})
  \urlprefix\url{https://doi.org/10.1063/1.1680919}

\bibitem{heinis+1993oms}
Heinis T, Chowdhury S and Kebarle P 1993 {\em Org.~Mass~Spectrom.\/} {\bf 28}
  358--365 (\textit{Preprint}
  \eprint{https://onlinelibrary.wiley.com/doi/pdf/10.1002/oms.1210280416})
  \urlprefix\url{https://onlinelibrary.wiley.com/doi/abs/10.1002/oms.1210280416}

\bibitem{ruoff+1995jpc}
Ruoff R~S, Kadish K~M, Boulas P and Chen E~C~M 1995 {\em J.~Phys.~C\/} {\bf 99}
  8843--8850 (\textit{Preprint} \eprint{https://doi.org/10.1021/j100021a060})
  \urlprefix\url{https://doi.org/10.1021/j100021a060}

\bibitem{naoto+2007jcp}
Ando N, Mitsui M and Nakajima A 2007 {\em J.~Chem.~Phys.\/} {\bf 127} 234305
  (\textit{Preprint} \eprint{https://doi.org/10.1063/1.2805185})
  \urlprefix\url{https://doi.org/10.1063/1.2805185}

\bibitem{hu+17jcc}
Hu Z, Zhou B, Sun Z and Sun H 2017 {\em J.~Comput.~Chem.\/} {\bf 38} 569--575
  (\textit{Preprint}
  \eprint{https://onlinelibrary.wiley.com/doi/pdf/10.1002/jcc.24736})
  \urlprefix\url{https://onlinelibrary.wiley.com/doi/abs/10.1002/jcc.24736}

\bibitem{zhu+18jpcc}
Zhu L, Yi Y and Wei Z 2018 {\em J.~Phys.~Chem.~C\/} {\bf 122} 22309--22316

\bibitem{pope-swen99book}
Pope M and Swenberg C~E 1999 {\em Electronic processes in organic crystals and
  polymers\/} vol~56 (Oxford University Press)

\bibitem{tamar+2009jacs}
Stein T, Kronik L and Baer R 2009 {\em J.~Am.~Chem.~Soc.\/} {\bf 131}
  2818--2820 pMID: 19239266 (\textit{Preprint}
  \eprint{https://doi.org/10.1021/ja8087482})
  \urlprefix\url{https://doi.org/10.1021/ja8087482}

\bibitem{jana1978prb}
Janak J~F 1978 {\em Phys.~Rev.~B\/} {\bf 18}(12) 7165--7168
  \urlprefix\url{https://link.aps.org/doi/10.1103/PhysRevB.18.7165}

\bibitem{zhen+17jpcl}
Zheng Z, Egger D~A, Bredas J~L, Kronik L and Coropceanu V 2017 {\em
  J.~Phys.~Chem.~Lett.\/} {\bf 8} 3277--3283

\bibitem{craciunescu+2022jpcl}
Craciunescu L, Wirsing S, Hammer S, Broch K, Dreuw A, Fantuzzi F, Sivanesan V,
  Tegeder P and Engels B 2022 {\em The Journal of Physical Chemistry Letters\/}
  {\bf 13} 3726--3731 pMID: 35442698 (\textit{Preprint}
  \eprint{https://doi.org/10.1021/acs.jpclett.2c00573})
  \urlprefix\url{https://doi.org/10.1021/acs.jpclett.2c00573}

\bibitem{jari+16nl}
Jariwala D, Howell S~L, Chen K~S, Kang J, Sangwan V~K, Filippone S~A, Turrisi
  R, Marks T~J, Lauhon L~J and Hersam M~C 2016 {\em Nano~Lett.\/} {\bf 16}
  497--503

\bibitem{liu+17nl}
Liu X, Gu J, Ding K, Fan D, Hu X, Tseng Y~W, Lee Y~H, Menon V and Forrest S~R
  2017 {\em Nano~Lett.\/} {\bf 17} 3176--3181

\bibitem{song+17nano}
Song Z, Schultz T, Ding Z, Lei B, Han C, Amsalem P, Lin T, Chi D, Wong S~L,
  Zheng Y~J, Li M~Y, Li L~J, Chen W, Koch N, Huang Y~L and Wee A~T~S 2017 {\em
  ACS~Nano\/} {\bf 11} 9128--9135

\bibitem{zhan+18am}
Zhang L, Sharma A, Zhu Y, Zhang Y, Wang B, Dong M, Nguyen H~T, Wang Z, Wen B,
  Cao Y, Liu B, Sun X, Yang J, Li Z, Kar A, Shi Y, Macdonald D, Yu Z, Wang X
  and Lu Y 2018 {\em Adv.~Mater.\/} {\bf 30} 1803986 (\textit{Preprint}
  \eprint{https://onlinelibrary.wiley.com/doi/pdf/10.1002/adma.201803986})
  \urlprefix\url{https://onlinelibrary.wiley.com/doi/abs/10.1002/adma.201803986}

\bibitem{gu+18acsp}
Gu J, Liu X, Lin E~c, Lee Y~H, Forrest S~R and Menon V~M 2018 {\em
  ACS~Photon.\/} {\bf 5} 100--104

\bibitem{amst+19nano}
Amsterdam S~H, Stanev T~K, Zhou Q, Lou A~J~T, Bergeron H, Darancet P, Hersam
  M~C, Stern N~P and Marks T~J 2019 {\em ACS~Nano\/} {\bf 13} 4183--4190

\bibitem{park+21as}
Park S, Mutz N, Kovalenko S~A, Schultz T, Shin D, Aljarb A, Li L~J, Tung V,
  Amsalem P, List-Kratochvil E~J, St\"ahler J, Xu X, Blumstengel S and Koch N
  2021 {\em Adv.~Sci.\/} {\bf 8} 2100215

\bibitem{abra1974}
Abramowitz M 1974 {\em Handbook of Mathematical Functions, With Formulas,
  Graphs, and Mathematical Tables\/} (USA: Dover Publications, Inc.) ISBN
  0486612724

\bibitem{temme1996}
Temme N~M 1996 {\em Special functions: an introduction to the classical
  functions of mathematical physics\/} (New York: Wiley)
  \urlprefix\url{http://site.ebrary.com/id/10452934}

\bibitem{kuma-taka1989prb}
Kumagai M and Takagahara T 1989 {\em Phys.~Rev.~B\/} {\bf 40}(18) 12359--12381
  \urlprefix\url{https://link.aps.org/doi/10.1103/PhysRevB.40.12359}

\bibitem{apsn1982ajp}
Aspnes D~E 1982 {\em Am.~J.~Phys.\/} {\bf 50} 704--709 (\textit{Preprint}
  \eprint{https://doi.org/10.1119/1.12734})
  \urlprefix\url{https://doi.org/10.1119/1.12734}

\bibitem{egui-hank1989prb}
Eguiluz A~G and Hanke W 1989 {\em Phys.~Rev.~B\/} {\bf 39}(14) 10433--10436
  \urlprefix\url{https://link.aps.org/doi/10.1103/PhysRevB.39.10433}

\bibitem{lang-kohn1973prb}
Lang N~D and Kohn W 1973 {\em Phys.~Rev.~B\/} {\bf 7}(8) 3541--3550
  \urlprefix\url{https://link.aps.org/doi/10.1103/PhysRevB.7.3541}

\bibitem{latu+2018npj}
Laturia A, Van~de Put M~L and Vandenberghe W~G 2018 {\em npj 2D Materials and
  Applications\/} {\bf 2} 6
  \urlprefix\url{https://doi.org/10.1038/s41699-018-0050-x}

\bibitem{chou+17jpcc}
Choudhury P, Ravavarapu L, Dekle R and Chowdhury S 2017 {\em
  J.~Phys.~Chem.~C\/} {\bf 121} 2959--2967

\bibitem{habi+20ats}
Habib M~R, Wang W, Khan A, Khan Y, Obaidulla S~M, Pi X and Xu M 2020 {\em
  Adv.~Theory~Simul.\/} {\bf 3} 2000045

\bibitem{aden2021jcp}
Adeniran O and Liu Z~F 2021 {\em J.~Chem.~Phys.\/} {\bf 155} 214702
  (\textit{Preprint} \eprint{https://doi.org/10.1063/5.0072995})
  \urlprefix\url{https://doi.org/10.1063/5.0072995}

\bibitem{mela+22pccp}
Melani G, Guerrero-Felipe J~P, Valencia A~M, Krumland J, Cocchi C and Iannuzzi
  M 2022 {\em Phys.~Chem.~Chem.~Phys.\/} {\bf 24}(27) 16671--16679
  \urlprefix\url{http://dx.doi.org/10.1039/D2CP01502A}

\bibitem{oliv+22prm}
Gonzalez~Oliva I, Caruso F, Pavone P and Draxl C 2022 {\em Phys. Rev.
  Materials\/} {\bf 6}(5) 054004
  \urlprefix\url{https://link.aps.org/doi/10.1103/PhysRevMaterials.6.054004}

\bibitem{guo+22nr}
Guo Y, Wu L, Deng J, Zhou L, Jiang W, Lu S, Huo D, Ji J, Bai Y, Lin X, Shunping
  Z, Xu H, Ji W and Zhang C 2022 {\em Nano~Res.\/} {\bf 15} 1276--1281

\bibitem{krum-cocc21es}
Krumland J and Cocchi C 2021 {\em Electron.~Struct.\/} {\bf 3} 044003

\bibitem{neaton+2006prl}
Neaton J~B, Hybertsen M~S and Louie S~G 2006 {\em Phys.~Rev.~Lett.\/} {\bf
  97}(21) 216405
  \urlprefix\url{https://link.aps.org/doi/10.1103/PhysRevLett.97.216405}

\bibitem{garc+2009prb}
Garcia-Lastra J~M, Rostgaard C, Rubio A and Thygesen K~S 2009 {\em
  Phys.~Rev.~B\/} {\bf 80}(24) 245427
  \urlprefix\url{https://link.aps.org/doi/10.1103/PhysRevB.80.245427}

\bibitem{sokl+2014apl}
Soklaski R, Liang Y and Yang L 2014 {\em Appl.~Phys.~Lett.\/} {\bf 104} 193110
  (\textit{Preprint} \eprint{https://doi.org/10.1063/1.4878098})
  \urlprefix\url{https://doi.org/10.1063/1.4878098}

\bibitem{huse+2013prb}
H\"user F, Olsen T and Thygesen K~S 2013 {\em Phys.~Rev.~B\/} {\bf 88}(24)
  245309 \urlprefix\url{https://link.aps.org/doi/10.1103/PhysRevB.88.245309}

\bibitem{qin+20jmcc}
Qin Z, Gao C, Wong W~W~H, Riede M~K, Wang T, Dong H, Zhen Y and Hu W 2020 {\em
  J. Mater. Chem. C\/} {\bf 8}(43) 14996--15008
  \urlprefix\url{http://dx.doi.org/10.1039/D0TC02746D}

\bibitem{ryou+2016sr}
Ryou J, Kim Y~S, KC S and Cho K 2016  {\bf 6} 29184 ISSN 2045-2322
  \urlprefix\url{https://doi.org/10.1038/srep29184}

\bibitem{vale+20pccp}
Valencia A~M, Guerrini M and Cocchi C 2020 {\em Phys. Chem. Chem. Phys.\/} {\bf
  22}(6) 3527--3538 \urlprefix\url{http://dx.doi.org/10.1039/C9CP06655A}

\bibitem{ping+10jpcl}
Pingel P, Zhu L, Park K~S, Vogel J~O, Janietz S, Kim E~G, Rabe J~P, Br\'{e}das
  J~L and Koch N 2010 {\em J.~Phys.~Chem.~Lett.\/} {\bf 1} 2037--2041

\bibitem{mend+13acie}
M\'{e}ndez H, Heimel G, Opitz A, Sauer K, Barkowski P, Oehzelt M, Soeda J,
  Okamoto T, Takeya J, Arlin J~B, Balandier J~Y, Geerts Y, Koch N and Salzmann
  I 2013 {\em Angew.~Chem.~Int.~Ed.\/} {\bf 52} 7751--7755
  \urlprefix\url{https://onlinelibrary.wiley.com/doi/abs/10.1002/anie.201302396}

\bibitem{goet+14jpcc}
Goetz K~P, Vermeulen D, Payne M~E, Kloc C, McNeil L~E and Jurchescu O~D 2014
  {\em J.~Phys.~Chem.~C\/} {\bf 2} 3065--3076

\bibitem{kief+19natm}
Kiefer D, Kroon R, Hofmann A~I, Sun H, Liu X, Giovannitti A, Stegerer D, Cano
  A, Hynynen J, Yu L, Zhang Y, Nai D, Harrelson T~F, Sommer M, Moulé A~J,
  Kemerink M, Marder S~R, McCulloch I, Fahlman M, Fabiano S and M\"{u}ller C
  2019 {\em Nature Materials\/} {\bf 18} 149-- 155

\bibitem{theu+21jpcc}
Theurer C~P, Valencia A~M, Hausch J, Zeiser C, Sivanesan V, Cocchi C, Tegeder P
  and Broch K 2021 {\em J.~Phys.~Chem.~C\/} {\bf 125} 6313--6323

\bibitem{russ+22jpcc}
Ru{\ss}egger N, Valencia A~M, Merten L, Zwadlo M, Duva G, Pithan L, Gerlach A,
  Hinderhofer A, Cocchi C and Schreiber F 2022 {\em J.~Phys.~Chem.~C\/} {\bf
  126} 4188--4198

\bibitem{sato+19jmcc}
Sato R, Kawamoto T and Mori T 2019 {\em J. Mater. Chem. C\/} {\bf 7}(3)
  567--577 \urlprefix\url{http://dx.doi.org/10.1039/C8TC05190A}

\bibitem{groz+08irpc}
Grozema F~C and Siebbeles L~D 2008 {\em Int.~Rev.~Phys.~Chem.\/} {\bf 27}
  87--138

\bibitem{troi11csr}
Troisi A 2011 {\em Chem.~Soc.~Rev.\/} {\bf 40} 2347--2358

\bibitem{ortm+11pssb}
Ortmann F, Bechstedt F and Hannewald K 2011 {\em Phys.~Status~Solidi~B\/} {\bf
  248} 511--525

\bibitem{mikh+15ees}
Mikhnenko O~V, Blom P~W and Nguyen T~Q 2015  {\bf 8} 1867--1888

\bibitem{enge-enge17pccp}
Engels B and Engel V 2017 {\em Phys.~Chem.~Chem.~Phys.\/} {\bf 19}(20)
  12604--12619

\bibitem{hoer+19cpc}
H\"ormann L, Jeindl A, Egger A~T, Scherbela M and Hofmann O~T 2019 {\em
  Computer Physics Communications\/} {\bf 244} 143--155 ISSN 0010-4655
  \urlprefix\url{https://www.sciencedirect.com/science/article/pii/S0010465519301973}

\bibitem{liu+22npjcm}
Liu X, Wang X, Gao S, Chang V, Tom R, Yu M, Ghiringhelli L~M and Marom N 2022
  {\em Npj~Comput.~Mater.\/} {\bf 8} 1--10

\bibitem{havu+09jcp}
Havu V, Blum V, Havu P and Scheffler M 2009 {\em J.~Comp.~Phys.\/} {\bf 228}
  8367 -- 8379 ISSN 0021-9991
  \urlprefix\url{http://www.sciencedirect.com/science/article/pii/S0021999109004458}

\bibitem{perd+96prl}
Perdew J~P, Burke K and Ernzerhof M 1996 {\em Phys.~Rev.~Lett.\/} {\bf 77}
  3865--3868

\bibitem{tkat-sche09prl}
Tkatchenko A and Scheffler M 2009 {\em Phys.~Rev.~Lett.\/} {\bf 102} 073005

\bibitem{brun12jcp}
Bruneval F 2012 {\em J.~Chem.~Phys.\/} {\bf 136} 194107

\bibitem{weig+02}
Weigend F, K\"{o}hn A and H\"{a}ttig C 2002 {\em J.~Chem.~Phys.\/} {\bf 116}
  3175--3183 (\textit{Preprint} \eprint{https://doi.org/10.1063/1.1445115})
  \urlprefix\url{https://doi.org/10.1063/1.1445115}

\bibitem{yana+04cpl}
Yanai T, Tew D~P and Handy N~C 2004 {\em Chem.~Phys.~Lett.\/} {\bf 393} 51 --
  57

\bibitem{geac-pope69jcp}
Geacintov N and Pope M 1969 {\em J.~Chem.~Phys.\/} {\bf 50} 814--822

\bibitem{tanc+2020jcp}
Tancogne-Dejean N, Oliveira M~J~T, Andrade X, Appel H, Borca C~H, Le~Breton G,
  Buchholz F, Castro A, Corni S, Correa A~A, De~Giovannini U, Delgado A, Eich
  F~G, Flick J, Gil G, Gomez A, Helbig N, Hübener H, Jestädt R, Jornet-Somoza
  J, Larsen A~H, Lebedeva I~V, Lüders M, Marques M~A~L, Ohlmann S~T, Pipolo S,
  Rampp M, Rozzi C~A, Strubbe D~A, Sato S~A, Schäfer C, Theophilou I, Welden A
  and Rubio A 2020 {\em J.~Chem.~Phys.\/} {\bf 152} 124119 (\textit{Preprint}
  \eprint{https://doi.org/10.1063/1.5142502})
  \urlprefix\url{https://doi.org/10.1063/1.5142502}

\bibitem{schl+gygi2015cpc}
Schlipf M and Gygi F 2015 {\em Comput.~Phys.~Commun.\/} {\bf 196} 36--44 ISSN
  0010-4655
  \urlprefix\url{https://www.sciencedirect.com/science/article/pii/S0010465515001897}

\bibitem{pasc+sill1990jcp}
Pascual-Ahuir J~L and Silla E 1990 {\em J.~Comput.~Chem.\/} {\bf 11} 1047--1060
  \urlprefix\url{https://onlinelibrary.wiley.com/doi/abs/10.1002/jcc.540110907}

\bibitem{gian+09jpcm}
Giannozzi P, Baroni S, Bonini N, Calandra M, Car R, Cavazzoni C, Ceresoli D,
  Chiarotti G~L, Cococcioni M, Dabo I, Corso A~D, de~Gironcoli S, Fabris S,
  Fratesi G, Gebauer R, Gerstmann U, Gougoussis C, Kokalj A, Lazzeri M,
  Martin-Samos L, Marzari N, Mauri F, Mazzarello R, Paolini S, Pasquarello A,
  Paulatto L, Sbraccia C, Scandolo S, Sclauzero G, Seitsonen A~P, Smogunov A,
  Umari P and Wentzcovitch R~M 2009 {\em J.~Phys.:~Condens.~Matter.\/} {\bf 21}
  395502

\bibitem{Sangalli_jpcm}
Sangalli D, Ferretti A, Miranda H, Attaccalite C, Marri I, Cannuccia E, Melo P,
  Marsili M, Paleari F, Marrazzo A, Prandini G, Bonf{\`{a}} P, Atambo M~O,
  Affinito F, Palummo M, Molina-S{\'{a}}nchez A, Hogan C, Gr\"{u}ning M,
  Varsano D and Marini A 2019 {\em J.~Phys.:~Condens.~Matter.\/} {\bf 31}
  325902

\bibitem{marz+12rmr}
Marzari N, Mostofi A~A, Yates J~R, Souza I and Vanderbilt D 2012 {\em
  Rev.~Mod.~Phys.\/} {\bf 84}(4) 1419--1475
  \urlprefix\url{https://link.aps.org/doi/10.1103/RevModPhys.84.1419}

\bibitem{pizz+2020jpcm}
Pizzi G, Vitale V, Arita R, Blügel S, Freimuth F, G{\'{e}}ranton G, Gibertini
  M, Gresch D, Johnson C, Koretsune T, Iba{\~{n}}ez-Azpiroz J, Lee H, Lihm J~M,
  Marchand D, Marrazzo A, Mokrousov Y, Mustafa J~I, Nohara Y, Nomura Y,
  Paulatto L, Ponc{\'{e}} S, Ponweiser T, Qiao J, Thöle F, Tsirkin S~S,
  Wierzbowska M, Marzari N, Vanderbilt D, Souza I, Mostofi A~A and Yates J~R
  2020 {\em J.~Phys.:~Condens.~Matter.\/} {\bf 32} 165902
  \urlprefix\url{https://doi.org/10.1088/1361-648x/ab51ff}

\bibitem{popescuZunger2012prb}
Popescu V and Zunger A 2012 {\em Phys.~Rev.~B\/} {\bf 85}(8) 085201
  \urlprefix\url{https://link.aps.org/doi/10.1103/PhysRevB.85.085201}

\bibitem{brun+gonz08prb}
Bruneval F and Gonze X 2008 {\em Phys.~Rev.~B\/} {\bf 78}(8) 085125
  \urlprefix\url{https://link.aps.org/doi/10.1103/PhysRevB.78.085125}

\bibitem{rozz+07prb}
Rozzi C~A, Varsano D, Marini A, Gross E~K~U and Rubio A 2006 {\em
  Phys.~Rev.~B\/} {\bf 73}(20) 205119
  \urlprefix\url{https://link.aps.org/doi/10.1103/PhysRevB.73.205119}

\bibitem{Hamann_prb}
Hamann D~R 2013 {\em Phys.~Rev.~B\/} {\bf 88}(8) 085117
  \urlprefix\url{https://link.aps.org/doi/10.1103/PhysRevB.88.085117}

\bibitem{Lanczos1950}
Lanczos C 1950 {\em J.~Res.~Natl.~Bur.~Stand.~B\/} {\bf 45} 255--282

\end{thebibliography}

\providecommand{\newblock}{}

\end{document}